\documentclass[aip,jcp,reprint,noshowkeys,superscriptaddress]{revtex4-1}
\usepackage{graphicx,dcolumn,bm,xcolor,microtype,multirow,amscd,amsmath,amssymb,amsfonts,physics,wrapfig,bbold,siunitx,xspace}
\usepackage[version=4]{mhchem}
\usepackage[normalem]{ulem}

\usepackage[utf8]{inputenc}
\usepackage[T1]{fontenc}

\usepackage{hyperref}
\hypersetup{
    colorlinks,
    linkcolor={red!50!black},
    citecolor={red!70!black},
    urlcolor={red!80!black}
}

\usepackage{listings}
\definecolor{codegreen}{rgb}{0.58,0.4,0.2}
\definecolor{codegray}{rgb}{0.5,0.5,0.5}
\definecolor{codepurple}{rgb}{0.25,0.35,0.55}
\definecolor{codeblue}{rgb}{0.30,0.60,0.8}
\definecolor{backcolour}{rgb}{0.98,0.98,0.98}
\definecolor{mygray}{rgb}{0.5,0.5,0.5}

\definecolor{sqred}{rgb}{0.85,0.1,0.1}
\definecolor{sqgreen}{rgb}{0.25,0.65,0.15}
\definecolor{sqorange}{rgb}{0.90,0.50,0.15}
\definecolor{sqblue}{rgb}{0.10,0.3,0.60}

\lstdefinestyle{mystyle}{
    backgroundcolor=\color{backcolour},
    commentstyle=\color{codegreen},
    keywordstyle=\color{codeblue},
    numberstyle=\tiny\color{codegray},
    stringstyle=\color{codepurple},
    basicstyle=\ttfamily\footnotesize,
    breakatwhitespace=false,
    breaklines=true,
    captionpos=b,
    keepspaces=true,
    numbers=left,
    numbersep=5pt,
    numberstyle=\ttfamily\tiny\color{mygray},
    showspaces=false,
    showstringspaces=false,
    showtabs=false,
    tabsize=2
  }

  \newcolumntype{d}{D{.}{.}{-1}}

\newcommand{\x}{$\cross$}

\newcommand{\mc}{\multicolumn}

\newcommand{\mcc}[1]{\multicolumn{1}{c}{#1}}

\newcommand{\ie}{\textit{i.e.}\xspace}
\newcommand{\eg}{\textit{e.g.}\xspace}
\newcommand{\etal}{\textit{et al.}\xspace}

\newcommand{\SupInf}{\textcolor{blue}{Supporting Information}\xspace}
\newcommand{\ph}{\phantom{(1)}}
\newcommand{\phh}{\phantom{(11)}}
\newcommand{\z}{\phantom{0}}

\newcommand{\LCPQ}{Laboratoire de Chimie et Physique Quantiques (UMR 5626), Universit\'e de Toulouse, CNRS, UPS, France}

\begin{document}	

\title{Reference Energies for Valence Ionizations and Satellite Transitions}
\author{Antoine \surname{Marie}}
	\email{amarie@irsamc.ups-tlse.fr}
	\affiliation{\LCPQ}
\author{Pierre-Fran\c{c}ois \surname{Loos}}
	\email{loos@irsamc.ups-tlse.fr}
	\affiliation{\LCPQ}
	
\begin{abstract}
Upon ionization of an atom or a molecule, another electron (or more) can be simultaneously excited.
These concurrently generated states are called ``satellites'' (or shake-up transitions) as they appear in ionization spectra as higher-energy peaks with weaker intensity and larger width than the main peaks associated with single-particle ionizations.
Satellites, which correspond to electronically excited states of the cationic species, are notoriously challenging to model using conventional single-reference methods due to their high excitation degree compared to the neutral reference state. 
This work reports 42 satellite transition energies and 58 valence ionization potentials of full configuration interaction (FCI) quality computed in small molecular systems. 
Following the protocol developed for the \textsc{quest} database [\href{https://doi.org/10.1002/wcms.1517}{V\'eril et al. \textit{Wiley Interdiscip.~Rev.: Comput.~Mol.~Sci.} \textbf{2021}, \textit{11}, e1517}], these reference energies are computed using the configuration interaction using a perturbative selection made iteratively (CIPSI) method.
In addition, the accuracy of the well-known coupled-cluster (CC) hierarchy (CC2, CCSD, CC3, CCSDT, CC4, and CCSDTQ) is gauged against these new accurate references.
The performances of various approximations based on many-body Green's functions ($GW$, GF2, and $T$-matrix) for ionization potentials are also analyzed. Their limitations in correctly modeling satellite transitions are discussed.
\bigskip
\begin{center}
       \boxed{\includegraphics[width=0.5\linewidth]{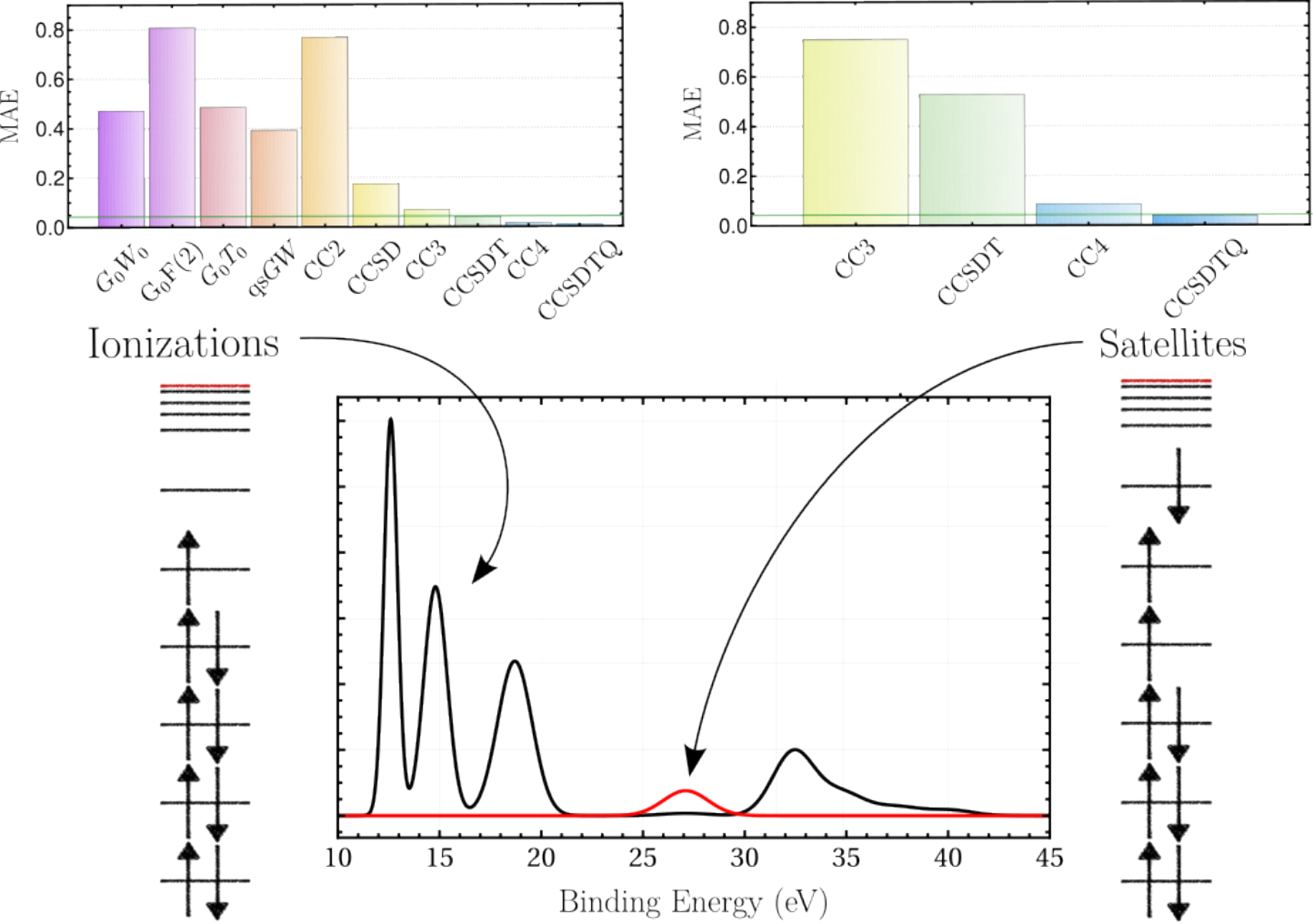}}
\end{center}
\end{abstract}

\maketitle

\section{Introduction}
\label{sec:introduction}

Ionization spectra, probed through techniques like UV-Vis, X-ray, synchrotron radiation, or electron impact spectroscopy, are invaluable tools in experimental chemistry for unraveling the structural intricacies of atoms, molecules, clusters, or solids. \cite{Carlson_1975,Hufner_2005,Fadley_2010} 
Through the positions and intensities of their peaks, these spectra offer key information about the sampled system.
For example, these measurements can be realized in various phases (gas, liquid, or solid) and, hence, analyzed to understand changes in electronic structure in these different phases. \cite{Campbell_1982,Winter_2004,Seidel_2016,Buttersack_2019}

Typically, within the energy range from \SIrange{10}{40}{\eV}, valence-shell ionization occurs, while the core shell is probed at significantly higher energies. \cite{Norman_2018}
This higher-energy region is not considered in the present study but the concepts that we discuss below in the context of valence-shell spectra are also encountered in the case of core electron spectroscopy.
Particularly, between \SI{10}{\eV} and \SI{20}{\eV}, ionization spectra of small molecules usually exhibit well-defined peaks.
These sharp and intense ionization peaks are essentially single-particle processes, \ie, an electron is ejected from the molecule and measured by the detector.
These first peaks are associated with outer-valence orbitals.
At slightly higher energies, typically several \si{\eV}, the situation is more complex as, in addition to inner-valence single-particle ionization peaks, additional broader and less intense peaks appear. 
These are referred to as satellites or shake-up transitions. 

In molecules, satellites represent ionization events coupled with the simultaneous excitation of one or more electrons.
They are thus intrinsically many-body phenomena, as one must describe at least two electrons and one hole. 
Satellite transitions can be seen as the equivalent of double excitations in the realm of neutral excitations. 
Because one must describe processes involving two electrons and two holes, double excitations pose significant challenges for theoretical methods, \cite{Starcke_2006,Loos_2019,doCasal_2023} and the same holds true for satellite transitions.
Consequently, such states can hardly be described by mean-field formalisms, such as Hartree-Fock (HF) theory. 
Thus, properly accounting for correlation effects is crucial to describe satellite transitions. \cite{Cederbaum_1974}
In particular, a recent study has emphasized the dynamic nature of this correlation. \cite{Mejuto-Zaera_2021}
In the following the term ``ionization'' is employed to refer to single-particle processes, also called Koopmans' states.

Theoretical benchmarks play a pivotal role in evaluating the accuracy of approximation methods. \cite{Pople_1989,Curtiss_1991,Curtiss_1998,Tajti_2004,Jureka_2006,Zhao_2006,Schreiber_2008,Goerigk_2010,Mardirossian_2017,Loos_2020d}
Concerning principal ionization potentials (IPs), which correspond to an electron detachment from the highest-occupied molecular orbital, two prominent benchmark sets are widely recognized: the extensively used $GW$100 test set \cite{vanSetten_2015,Caruso_2016,Krause_2017} and a set comprising 24 organic acceptor molecules. \cite{Gallandi_2016,Richard_2016,Knight_2016,Dolgounitcheva_2016}
Both sets rely on reference values obtained from coupled-cluster (CC) with singles, doubles, and perturbative triples [CCSD(T)] calculations \cite{Cizek_1966,Purvis_1982,Raghavachari_1989} and determined by the energy difference between the neutral and cationic ground-state energies. \cite{Shavitt_2009}
Recently, Ranasinghe \etal~created a comprehensive benchmark set including not only principal IPs but also outer- and inner-valence IPs of organic molecules. \cite{Ranasinghe_2019}
Reference values for this set were computed using the IP version of the equation-of-motion (IP-EOM) formalism \cite{Stanton_1993a,Watts_1994,Musial_2003a,Kamiya_2006,Gour_2006} of CC theory with up to quadruple excitations (IP-EOM-CCSDTQ) for the smallest molecules. \cite{Ranasinghe_2019}
Note that these benchmarks and the present work deal with vertical IPs, and we shall not address their adiabatic counterparts here.

To demonstrate its predictive capability for valence ionization spectra, an electronic structure method must precisely locate the positions of both outer- and inner-valence ionization potentials, along with valence satellites.
However, to the best of our knowledge, no established theoretical benchmarks exist for satellite energies in molecules. 
Consequently, the primary goal of this manuscript is to establish such a set of values.
Finally, it is important to mention that to be fully predictive a method should be able to predict the intensities associated with of these transitions.
However, benchmarking intensities is beyond the scope of this work and will be considered in a future study.

Nowadays, a plethora of methods exist to compute IPs in molecular systems.
The most straightforward among them is HF where occupied orbital energies serve as approximations of the IPs (up to a minus sign) by virtue of Koopmans' theorem. \cite{SzaboBook}
Similarly, within density-functional theory (DFT), the Kohn-Sham (KS) orbital energies can be used as approximate ionization energies. \cite{Perdew_1982,Ivanov_1999,Chong_2002}
The accurate computation of IPs within KS-DFT is still an ongoing research field with, for example, long-range corrected functionals, \cite{Hirao_2022} KS potential adjustors, \cite{Gorling_2015,Thierbach_2017} double-hybrids functionals, \cite{Mester_2023} or even functionals directly optimized for IPs. \cite{Verma_2014,Jin_2016,Jin_2018,Ranasinghe_2019}
An alternative way to compute electron detachment energies at the HF or KS-DFT levels is through the state-specific self-consistent-field ($\Delta$SCF) formalism, where one optimizes both the neutral ground state and the cationic state of interest, the IPs being computed as the difference between these two total energies.
This strategy has been mainly used to compute core binding energies and is known to perform better than Koopmans' theorem thanks to orbital relaxation. \cite{Bagus_1965,Guest_1975,Besley_2009,PueyoBellafont_2016,Jorstad_2022,Kahk_2023,Hirao_2023} 

Mean-field methods, such as HF and KS-DFT, provide a first approximation to IPs but greater accuracy is often required.
The well-known configuration interaction (CI) and CC formalisms provide two systematically improvable paths toward the exact IPs. \cite{Olsen_1996a,Musial_2003a,Kamiya_2006,Gour_2006}
Within both frameworks, IPs can be obtained through a diagonalization of a given Hamiltonian matrix in the $(N-1)$-electron sector of the Fock space or through a state-specific formalism similar to $\Delta$SCF. 
Ranasinghe \etal~have shown that the mean absolute error (MAE) of IP-EOM-CCSDT with respect to CCSDTQ is only \SI{0.03}{\eV} for a set containing 42 IPs of small molecules. \cite{Ranasinghe_2019}
Considering the same set, the cheaper IP-EOM-CCSD method has a MAE of \SI{0.2}{\eV}.
Recently, the unitary CC formalism has also been employed within the IP-EOM formalism to compute IPs. \cite{Dempwolff_2022}
As mentioned above, the $\Delta$SCF strategy can be extended to correlated methods which leads to the $\Delta$CC method as an alternative to obtain IPs. \cite{Krause_2015} 
Once again, it has been mainly used to compute core IPs but it is also possible to determine valence IPs. \cite{Zheng_2019,Lee_2019,Zheng_2022}
Selected CI (SCI)  \cite{Bender_1969,Whitten_1969,Huron_1973,Buenker_1974} provides yet another systematically improvable formalism for IPs.  
Indeed, by increasing progressively the number of determinants included in the variational space, one can in principle reach any desired accuracy, up to the full CI (FCI) limit. \cite{Eriksen_2020,Eriksen_2021,Caffarel_2016,Schriber_2016,Holmes_2017,Damour_2021,Larsson_2022}
Recently, the adaptative sampling CI algorithm \cite{Schriber_2016,Tubman_2016,Tubman_2018,Tubman_2020} has been used to compute accurate valence ionization spectra of small molecules. \cite{Mejuto-Zaera_2021}

In contrast to the wave function methods previously mentioned, one can also compute IPs via a more natural way based on electron propagators (or Green's functions), such as the $GW$ approximation \cite{Martin_2016,Golze_2019,Marie_2023a} or the algebraic diagrammatic construction (ADC). \cite{Schirmer_2018,Banerjee_2023}
The $GW$ methodology has a myriad of variants. Its one-shot $G_0W_0$ version, \cite{Strinati_1980,Hybertsen_1985a,Godby_1988,Linden_1988,Northrup_1991,Blase_1994,Rohlfing_1995} which was first popularized in condensed matter physics, is now routinely employed to compute IPs of molecular systems and can be applied to systems with thousands of correlated electrons. \cite{Neuhauser_2013,Neuhauser_2014,Govoni_2015,Vlcek_2017,Wilhelm_2018,Duchemin_2019,DelBen_2019,Forster_2020,Forster_2021,Duchemin_2020,Duchemin_2021,Forster_2022,Panades-Barrueta_2023} 
Other flavors of $GW$ such as eigenvalue-only self-consistent $GW$ (ev$GW$) \cite{Shishkin_2007a,Blase_2011b,Marom_2012,Wilhelm_2016,Kaplan_2016} and quasi-particle self-consistent $GW$ (qs$GW$) \cite{Kaplan_2016,Faleev_2004,vanSchilfgaarde_2006,Kotani_2007,Ke_2011,Marie_2023} have also been benchmarked for IPs.
Although the $GW$ method is by far the most popular approach nowadays, there exist some alternatives, such as the second Born [also known as second-order Green's function (GF2) in the quantum chemistry community] \cite{Casida_1989,Casida_1991,SzaboBook,Stefanucci_2013,Ortiz_2013,Phillips_2014,Phillips_2015,Rusakov_2014,Rusakov_2016,Hirata_2015,Hirata_2017,Backhouse_2021,Backhouse_2020b,Backhouse_2020a,Pokhilko_2021a,Pokhilko_2021b,Pokhilko_2022} or the $T$-matrix\cite{Liebsch_1981,Bickers_1989a,Bickers_1991,Katsnelson_1999,Katsnelson_2002,Zhukov_2005,vonFriesen_2010,Romaniello_2012,Gukelberger_2015,Muller_2019,Friedrich_2019,Biswas_2021,Zhang_2017,Li_2021b,Li_2023,Loos_2022a,Orlando_2023a,Orlando_2023b} approximations. However, none of them has enjoyed the popularity and performances reached by $GW$. \cite{Lewis_2019,Bruneval_2021,Monino_2023}
On the darker side, one of the main flaws of the $GW$ approximation is its lack of systematic improvability, especially compared to the wave function methods mentioned above.
Various beyond-$GW$ schemes have been designed and gauged, but none of them seem to offer, at a reasonable cost, a systematic route toward exact IPs. \cite{Baym_1961,Baym_1962,DeDominicis_1964a,DeDominicis_1964b,Bickers_1989a,Bickers_1989b,Bickers_1991,Hedin_1999,Bickers_2004,Shirley_1996,DelSol_1994,Schindlmayr_1998,Morris_2007,Shishkin_2007b,Romaniello_2009a,Romaniello_2012,Gruneis_2014,Hung_2017,Maggio_2017b,Mejuto-Zaera_2022,Wen_2024}

The prediction of satellite peaks in molecules garnered attention in the late 20th century.
In the 70's, Schirmer and coworkers applied extensively the 2ph-TDA [and the closely related ADC(3)] formalism to study the inner-valence region of small molecules. \cite{Cederbaum_1974,Schirmer_1977,Cederbaum_1977,Schirmer_1978,Schirmer_1978a,Schirmer_1978b,Domcke_1978,Cederbaum_1978,Cederbaum_1980,vonNiessen_1980,vonNiessen_1981,Walter_1981,Schirmer_1983a,Cederbaum_1986}
CI methods were also employed by other groups to study this energetic region. \cite{Bagus_1977,Kosugi_1979,Kosugi_1981,Honjou_1981,Nakatsuji_1982,Arneberg_1982,Roy_1986,Bawagan_1988,Bawagan_1988a,Clark_1989,Clark_1990,Lisini_1991}
In both formalisms, the satellite energies are easily accessible as they correspond to higher-energy roots of the ADC and CI matrices. 
After relative successes for outer-valence ionizations, it was quickly realized that the inner-valence shell is much more difficult to describe due to the overlap between the inner-valence ionization and the outer-valence satellite peaks. \cite{Schirmer_1978b}
As mentioned above, the satellites present in this energy range cannot be described without taking into account electron correlation at a high level of theory.
Even more troublesome, in some cases, the orbital picture (or quasiparticle approximation) completely breaks down.
In other words, it becomes meaningless to assign the character of ionization or satellite to a given transition. \cite{Schirmer_1977,Schirmer_1978b}
In the following decades, the symmetry-adapted-cluster (SAC) CI was extensively used to study the inner-valence ionization spectra of small organic molecules. \cite{Nakatsuji_1983,Wasada_1989,Nakatsuji_1991,Ehara_1998,Ehara_1999,Ehara_1999a,Ehara_2001,Ishida_2002,Ehara_2003,Ohtsuka_2006,Ning_2008,Tian_2012}
SAC-CI was shown to be able to compute satellite energies in quantitative agreement with experiments while methods based on Green's functions have been in qualitative agreement, at best.

Satellites, sometimes called sidebands, have been extensively studied in the context of materials. \cite{Martin_2016}
These additional peaks, which can have different natures, are observed in photoemission spectra of metals, semiconductors, and insulators. \cite{Aryasetiawan_1996,Vos_2001,Vos_2002,Kheifets_2003,Guzzo_2011,Guzzo_2014,Lischner_2013,Zhou_2015,Vigil-Fowler_2016,Vlcek_2018} In ``simple'' metals, such as bulk sodium \cite{Aryasetiawan_1996,Zhou_2015} or its paradigmatic version, the uniform electron gas, \cite{Holm_1997,Holm_2000,Lischner_2014,Kas_2014,McClain_2016,Vigil-Fowler_2016,Caruso_2016,Kas_2022a} satellites are usually created by the strong coupling between electrons and plasmon excitations. It is widely recognized that $GW$ does not properly describe satellite structures in solids, and it is required to include vertex corrections to describe these many-body effects. One of the most common schemes to study satellites in solids is the cumulant expansion, \cite{Guzzo_2011,Kas_2014,Mayers_2016,Zhou_2018,Tzavala_2020}
which is formally linked to electron-boson Hamiltonians. \cite{Lundqvist_1969,Langreth_1970,Hedin_1980,Hedin_1999}

Nowadays, computational and theoretical progress allows us to systematically converge to exact neutral excitation energies of small molecules, \cite{Holmes_2017,Garniron_2018,Chien_2018,Loos_2018a,Loos_2019,Loos_2020c} and this holds as well for charged excitations like IPs. 
For example, Olsen \etal~computed the exact first three IPs of water using FCI, \cite{Olsen_1996a} while Kamiya and Hirata went up to IP-EOM-CCSDTQ to compute highly accurate satellite energies for \ce{CO} and \ce{N2}. \cite{Kamiya_2006}
As mentioned previously, a set of 42 IPs of CCSDTQ quality is also available now. \cite{Ranasinghe_2019}
Finally, Chatterjee and Sokolov recently computed 27 valence IPs using the semistochastic heatbath SCI method \cite{Holmes_2016,Holmes_2017,Sharma_2017} in order to benchmark their multi-reference implementation of ADC. \cite{Chatterjee_2019,Chatterjee_2020}
They also report FCI-quality energies for the four lowest satellite states of the carbon dimer.
The present manuscript contributes to this line of research by providing 42 satellite energies of FCI quality.
Additionally, 58 valence IPs are reported as well, among which 37 were not present in Ranasinghe's CCSDTQ nor Chatterjee's FCI benchmark set. \cite{Ranasinghe_2019,Chatterjee_2019}
This study is part of a larger database of highly accurate vertical neutral excitation energies named \textsc{quest} which now includes more than 900 excitation energies. \cite{Loos_2018a,Loos_2019,Loos_2020c,Loos_2020d,Loos_2020f,Veril_2021,Loos_2021b,Loos_2022b,Jacquemin_2023,Loos_2023,Loos_2024}
Our hope is that these new data will serve as a valuable resource for encouraging the development of novel approximate methods dedicated to computing satellite energies, building on the success of benchmarks with highly-accurate reference energies and properties.

\section{Computational details}
\label{sec:comp_det}

The geometries of the molecular systems considered here have been optimized using \textsc{cfour} \cite{CFOUR} following \textsc{quest}'s protocol, \cite{Loos_2020d,Veril_2021} \textit{i.e.} at the CC3/aug-cc-pVTZ level \cite{Christiansen_1995b,Koch_1997} without frozen-core approximation.
The corresponding cartesian coordinates can be found in the \SupInf.
Throughout the paper the basis sets considered are Pople's 6-31+G* \cite{Gordon_1982,Francl_1982,Clark_1983,Ditchfield_2003,Hehre_2003,Dill_2008,Binkley_2008} and Dunning's aug-cc-pVXZ (where X = D, T, and Q). \cite{Dunning_1989,Kendall_1992,Prascher_2011,Woon_1993}

\subsection{Selected CI calculations}

All SCI calculations have been performed using the configuration interaction using a perturbative selection made iteratively (CIPSI) algorithm \cite{Huron_1973,Giner_2013,Giner_2015,Caffarel_2016b,Garniron_2017,Garniron_2018} as implemented in \textsc{quantum package}. \cite{Garniron_2019}
The multi-state CIPSI calculations are performed by converging several eigenvalues (using the iterative Davidson diagonalization procedure) at each iteration and then selecting determinants for these eigenvectors in a state-averaged way. \cite{Scemama_2019}
For more details about the CIPSI method and its implementation, see Ref.~\onlinecite{Garniron_2019}.
The frozen-core approximation has been enforced in all calculations using the conventions of \textsc{gaussian16} \cite{g16} and \textsc{cfour}, \cite{CFOUR} except for \ce{Li} and \ce{Be} where the 1s orbital was not frozen.

We followed a two-step procedure to obtain the ionization and satellite energies, $I_{\nu}^{N}$, at the SCI level.
First, two single-state calculations are performed for the $N$- and $(N-1)$-electron ground states. This yields the principal IP of the system, $I_{0}^{N} = E_0^{N-1} - E_0^{N}$, where $E_0^{N-1}$ and $E_0^{N}$ are the ground-state energies of the $N$- and $(N-1)$-electron systems, respectively.
Then, a third, multi-state calculation is performed to compute the neutral excitation energies of the $(N-1)$-electron system, $\Delta E_\nu^{N-1} = E_\nu^{N-1} - E_0^{N-1}$, where $E_\nu^{N-1}$ is the energy of the $\nu$th excited states associated with the $(N-1)$-electron system. Combining these three calculations, one gets
\begin{equation}
  \label{eq:E_ip}
  I_\nu^{N}  = E_0^{N-1} - E_0^{N} + \Delta E_\nu^{N-1}
\end{equation}
Because single-state calculations converge faster than their multi-state counterparts, the limiting factor associated with the present CIPSI calculations are the convergence of the excitation energies $\Delta E_\nu^{N-1}$, and this is what determines ultimately the overall accuracy of $I_\nu^{N}$.

For each system and state, the SCI variational energy has been extrapolated as a function of the second-order perturbative correction using a linear weighted fit using the last 3 to 6 CIPSI iterations. \cite{Holmes_2017,Damour_2021,Damour_2023,Burton_2023} The weights have been taken as the square of the inverse of the perturbative correction. 
The estimated FCI energy is then chosen amongst these extrapolated values obtained with a variable number of points such that the standard error associated with the extrapolated energy is minimal.
Below, we report error bars associated with these extrapolated FCI values. However, it is worth remembering that these do not correspond to genuine statistical errors.
The fitting procedure has been performed with \textsc{mathematica} using default settings. \cite{Mathematica}

The SCI values and their corresponding error bars are reported with three decimal places to enable fair and reliable comparisons between methods, ensuring a precision well below the chemical accuracy threshold (\ie \SI{0.043}{\eV}).
One should keep in mind that only the first decimal might be experimentally meaningful (\ie, measured without uncertainty).
Finally, note that comparing theoretical and experimental IPs is a complex task, requiring consideration of vibrational effects (see for example Ref.~\onlinecite{Alvertis_2024}) and possibly relativistic effects for inner-valence IP and/or molecules containing third-row atoms.

\subsection{Coupled-cluster calculations}
\label{sec:CC}

The EOM-CC calculations have been done using \textsc{cfour} with the default convergence thresholds. \cite{CFOUR} Again, the frozen-core approximation was enforced systematically.
IP-EOM-CC calculations, \ie, diagonalization of the CC effective Hamiltonian in the $(N-1)$-electron sector of the Fock space, \cite{Stanton_1993a,Watts_1994,Musial_2003a,Kamiya_2006,Gour_2006} have been performed for CCSD, \cite{Purvis_1982,Scuseria_1987,Koch_1990a,Koch_1990c,Stanton_1993a,Stanton_1993b} CCSDT, \cite{Noga_1987,Scuseria_1988,Watts_1994,Kucharski_2001} and CCSDTQ. \cite{Kucharski_1991,Kallay_2001,Hirata_2004,Kallay_2003,Kallay_2004a} 
At the CCSD level, the EOM space includes the one-hole (1h) and the two-hole-one-particle (2h1p) configurations, while the three-hole-two-particle (3h2p) and four-hole-three-particle (4h3p) configurations are further added at the CCSDT and CCSDTQ, respectively. Note that, within the CC formalism, we assume that the IP and electron affinity (EA) sectors are decoupled. \cite{Nooijen_1995,Rishi_2020,Quintero_2022,Tolle_2023}
For CC2, \cite{Christiansen_1995a,Hattig_2000} CC3, \cite{Christiansen_1995b,Koch_1995,Koch_1997,Hald_2001,Paul_2021} and CC4, \cite{Kallay_2004b,Kallay_2005,Loos_2021a,Loos_2022b} diagonalization in the $(N-1)$-electron sector of the Fock space is not available yet.
Hence, it has been carried out in the $N$-electron sector of the Fock space \cite{Rowe_1968a,Emrich_1981,Sekino_1984,Geertsen_1989,Stanton_1993a,Comeau_1993,Watts_1994} with an additional very diffuse (or bath) orbital with zero energy to obtain ionization and satellite energies.
Therefore, at the CC2 level, the EOM space includes the one-hole-one-particle (1h1p) and the two-hole-two-particle (2h2p) configurations, while the three-hole-three-particle (3h3p) and four-hole-four-particle (4h4p) configurations are further added at the CC3 and CC4 levels, respectively.
These two schemes produce identical IPs and satellite energies but, for a given level of theory, the diagonalization in the $N$-electron sector is more computationally demanding due to the larger size of the EOM space (see Ref.~\onlinecite{Stanton_1999} for more details).
In each scheme, the desired states have been obtained thanks to the root-following Davidson algorithm implemented in \textsc{cfour}.
The initial vectors were built using the dominant configurations of the SCI vectors.

The $\Delta$CCSD(T) calculations have been performed with \textsc{gaussian16}. \cite{g16}
These calculations are based on a closed-shell restricted HF reference and an open-shell unrestricted HF reference for the neutral and cationic species, respectively. \cite{Bruneval_2021}

\subsection{Green's function calculations}

Many-body Green's function calculations have been carried out with the open-source software \textsc{quack}. \cite{QuAcK}
In the following, we use the acronyms $G_0W_0$, G$_0$F(2), and $G_0T_0$ to refer to the one-shot schemes where one relies on the $GW$, second-Born, and $T$-matrix self-energies, respectively.
Each approximated scheme considered in this work ($G_0W_0$, qs$GW$, G$_0$F(2), and $G_0T_0$) relies on HF quantities as starting point.
We refer the reader to Refs.~\onlinecite{Monino_2023,Marie_2023a} for additional details about the theory and implementation of these methods.
The infinitesimal broadening parameter $\eta$ is set to \SI{0.001}{\hartree} for all calculations.
It is worth mentioning that we do not linearize the quasiparticle equation to obtain the quasiparticle energies.
The qs$GW$ calculations are performed with the regularized scheme based on the similarity renormalization group approach, as described in Ref.~\onlinecite{Marie_2023}. A flow parameter of $s = 500$ is employed. All (occupied and virtual) orbitals are corrected.
The spectral weight of each quasiparticle solution is reported in \SupInf.
Compared to the EOM-CC formalism discussed in Sec.~\ref{sec:CC}, it is important to mention that, in the Green's function framework, the IP and EA sectors [\ie~the 1h and one-particle (1p) configurations] are actually coupled, \cite{Lange_2018,Quintero_2022,Monino_2023} effectively creating higher-order diagrams. \cite{Lange_2018,Schirmer_2018}

\section{Results and Discussion}
\label{sec:discussion}

The present section is partitioned into subsections, each dedicated to a distinct group of related molecules.
Within these subsections, we focus our attention mainly on the satellite states while IPs are addressed in Sec.~\ref{sec:stat}.

Each state considered in this work is reported alongside its symmetry label, \eg~$1~^{2}\mathrm{B}_1$ for the principal IP of water.
Furthermore, the main orbitals involved in the ionization process are specified.
For example, the $N$-electron ground state of water has the following dominant configuration $(1a_1)^2(2a_1)^2(1b_2)^2(3a_1)^2(1b_1)^2$, while the configuration of the $(N-1)$-electron ground state is $(1a_1)^2(2a_1)^2(1b_2)^2(3a_1)^2(1b_1)^1$.
Hence, we denote the principal IP as $(1b_1)^{-1}$ to indicate that an electron has been ionized from the $1b_1$ orbital.
The lowest satellite of water, \ie~the $2~^{2}\mathrm{B}_1$ state of configuration $(1a_1)^2(2a_1)^2(1b_2)^2(3a_1)^1(1b_1)^1(4a_1)^1$, is labeled as $(3a_1)^{-1}(1b_1)^{-1}(4a_1)^1$ to signify that one electron was detached from the orbital $1b_1$ and $3a_1$, one of them being subsequently promoted to the virtual orbital $4a_1$ and the other ionized.
In some cases, additional valence complete-active-space CI calculations have been performed using \textsc{molpro} to determine the symmetry of the FCI states. \cite{Molpro}


\begin{squeezetable}
\begin{table*}
  \caption{Valence ionizations and satellite transition energies (in \si{\eV}) of the 10-electron series for various methods and basis sets. 
  AVXZ stands for aug-cc-pVXZ (where X = D, T, and Q). Selected experimental values are also reported.}
  \label{tab:tab1}
  \begin{ruledtabular}
    \begin{tabular}{rccccccccccccc}
      & \mc{4}{c}{Basis} & \mc{4}{c}{Basis} & \mc{4}{c}{Basis} \\
      \cline{2-5} \cline{6-9} \cline{10-13}
      Methods & \mcc{6-31$+$G$^{*}$} & \mcc{AVDZ} & \mcc{AVTZ} & \mcc{AVQZ} & \mcc{6-31$+$G$^{*}$} & \mcc{AVDZ} & \mcc{AVTZ} & \mcc{AVQZ} & \mcc{6-31$+$G$^{*}$} & \mcc{AVDZ} & \mcc{AVTZ} & \mcc{AVQZ}  \\
      \hline
      Mol .       & \mc{12}{c}{Water (\ce{H2O})} \\
      State/Conf. & \mc{4}{c}{$1~^{2}\mathrm{B}_1$/$(1b_1)^{-1}$} & \mc{4}{c}{$1~^{2}\mathrm{A}_1$/$(3a_1)^{-1}$} & \mc{4}{c}{$1~^{2}\mathrm{B}_2$/$(1b_2)^{-1}$} \\
      Exp.        &  \mc{4}{c}{12.6 \cite{Ning_2008}} & \mc{4}{c}{14.8 \cite{Ning_2008}} & \mc{4}{c}{18.7 \cite{Ning_2008}} \\
      \cline{2-5} \cline{6-9} \cline{10-13}
      CC2         & 11.159 & 11.345 & 11.541 & 11.620 & 13.513 & 13.645 & 13.791 & 13.863 & 18.035 & 18.039 & 18.145 & 18.211 \\
      CCSD        & 12.170 & 12.386 & 12.594 & 12.675 & 14.502 & 14.677 & 14.825 & 14.895 & 18.861 & 18.888 & 18.972 & 19.032 \\
      CC3         & 12.287 & 12.519 & 12.661 & 12.722 & 14.621 & 14.811 & 14.899 & 14.949 & 18.950 & 18.993 & 19.023 & 19.065 \\
      CCSDT       & 12.276 & 12.491 & 12.629 & 12.689 & 14.601 & 14.776 & 14.861 & 14.910 & 18.919 & 18.951 & 18.981 & 19.022 \\
      CC4         & 12.307 & 12.543 & 12.683 & 12.741 & 14.635 & 14.832 & 14.920 & 14.968 & 18.952 & 18.999 & 19.030 & 19.070 \\
      CCSDTQ      & 12.304 & 12.534 & 12.673 &     \x & 14.631 & 14.822 & 14.907 &     \x & 18.947 & 18.990 & 19.018 &     \x \\
      FCI         & 12.309 & 12.540 & 12.679 & 12.737 & 14.636 & 14.829 & 14.915 & 14.962 & 18.950 & 18.995 & 19.024 & 19.063 \\
      $G_0W_0$    & 12.312 & 12.485 & 12.884 & 13.080 & 14.625 & 14.781 & 15.106 & 15.285 & 18.818 & 18.865 & 19.129 & 19.290 \\
      qs$GW$      & 12.379 & 12.640 & 12.879 & 12.982 & 14.696 & 14.932 & 15.107 & 15.197 & 18.965 & 19.069 & 19.188 & 19.271 \\
      G$_0$F(2)   & 11.110 & 11.279 & 11.555 & 11.675 & 13.507 & 13.626 & 13.837 & 13.945 & 17.983 & 17.978 & 18.141 & 18.236 \\
      $G_0T_0$    & 11.967 & 12.095 & 12.357 &     \x & 14.240 & 14.336 & 14.532 &     \x & 18.459 & 18.429 & 18.572 &     \x \\
      \hline
      Mol.        & \mc{12}{c}{Ammonia (\ce{NH3})} \\
      State/Conf. & \mc{4}{c}{$1~^{2}\mathrm{A}_1$/$(3a_1)^{-1}$} & \mc{4}{c}{$1~^{2}\mathrm{E}$/$(1e_g)^{-1}$} & \mc{4}{c}{$3~^{2}\mathrm{A}_1$/$(2a_1)^{-1}$} \\
      Exp.        & \mc{4}{c}{10.93\cite{Edvardsson_1999}}& \mc{4}{c}{16.6\cite{Edvardsson_1999}} & \mc{4}{c}{} \\
      \cline{2-5} \cline{6-9} \cline{10-13}
      CC2         & \z9.779 & \z9.986 & 10.168 & 10.234 & 15.794 & 15.828 & 15.960 & 16.019 & 27.646 & 27.381 & 27.365 &     \x\ph \\
      CCSD        &  10.434 &  10.677 & 10.862 & 10.923 & 16.403 & 16.473 & 16.588 & 16.639 & 27.745 & 27.696 & 27.855 & 27.915\ph \\
      CC3         &  10.447 &  10.746 & 10.888 & 10.935 & 16.407 & 16.520 & 16.592 & 16.629 & 27.252 & 27.114 & 27.191 &     \x\ph \\
      CCSDT       &  10.449 &  10.734 & 10.876 & 10.922 & 16.399 & 16.500 & 16.573 & 16.609 & 26.773 & 26.724 & 26.899 &     \x\ph \\
      CC4         &  10.461 &  10.761 & 10.901 &     \x & 16.417 & 16.533 & 16.603 &     \x & 26.669 & 26.621 & 26.746 &     \x\ph \\
      CCSDTQ      &  10.461 &  10.760 & 10.899 &     \x & 16.415 & 16.529 & 16.598 &     \x & 26.698 & 26.645 & 26.768 &     \x\ph \\
      FCI         &  10.463 &  10.762 & 10.901 & 10.945 & 16.418 & 16.534 & 16.603 & 16.640 & 26.683 & 26.659 & 26.779 & 26.833(1) \\
      $G_0W_0$    &  10.675 &  10.837 & 11.201 & 11.362 & 16.527 & 16.578 & 16.867 & 17.007 & 28.241 & 28.117 & 28.427 & 28.463\ph \\
      qs$GW$      &  10.520 &  10.870 & 11.094 & 11.176 & 16.468 & 16.655 & 16.805 & 16.878 & 28.029 & 27.962 & 27.980 & 28.151\ph \\
      G$_0$F(2)   & \z9.841 & \z9.994 & 10.244 & 10.345 & 15.817 & 15.814 & 16.002 & 16.088 & 27.589 & 27.638 & 27.729 &     \x\ph \\
      $G_0T_0$    &  10.399 &  10.497 & 10.716 &     \x & 16.217 & 16.170 & 16.330 &     \x & 28.860 & 28.738 & 28.860 &     \x\ph \\
      \hline
      Mol.        & \mc{8}{c}{Methane (\ce{CH4})} & \mc{4}{c}{Hydrogen fluoride (\ce{HF})} \\
      State/Conf. & \mc{4}{c}{$1~^{2}\mathrm{T}_2$/$(1t_2)^{-1}$}  & \mc{4}{c}{$1~^{2}\mathrm{A}_1$/$(2a_1)^{-1}$} & \mc{4}{c}{$1~^{2}\mathrm{\Pi}$/$(1\pi)^{-1}$} \\
      Exp.        & \mc{4}{c}{14.5\cite{Gothe_1991}} & \mc{4}{c}{23.0\cite{Gothe_1991}} & \mc{4}{c}{16.19 \cite{Bieri_1980}} \\
      \cline{2-5} \cline{6-9} \cline{10-13}
      CC2         & 13.787 & 13.888 & 14.028 & 14.079 & 23.289 & 23.227 & 23.311 & 23.352\phh & 14.431 & 14.559 & 14.725 & 14.813 \\
      CCSD        & 14.102 & 14.258 & 14.387 & 14.428 & 23.238 & 23.247 & 23.383 & 23.426\phh & 15.688 & 15.837 & 16.021 & 16.117 \\
      CC3         & 14.060 & 14.270 & 14.365 & 14.395 & 23.034 & 23.035 & 23.138 & 23.173\phh & 15.917 & 16.036 & 16.126 & 16.194 \\
      CCSDT       & 14.068 & 14.269 & 14.365 & 14.395 & 23.040 & 23.035 & 23.135 & 23.171\phh & 15.885 & 15.992 & 16.077 & 16.145 \\
      CC4         & 14.072 & 14.284 & 14.376 &     \x & 23.039 & 23.050 & 23.142 &     \x\phh & 15.947 & 16.068 & 16.161 & 16.227 \\
      CCSDTQ      & 14.073 & 14.284 & 14.376 &     \x & 23.042 & 23.052 & 23.143 &     \x\phh & 15.935 & 16.051 & 16.140 & 16.205 \\
      FCI         & 14.073 & 14.285 & 14.377 & 14.407 & 23.043 & 23.056 & 23.146 & 23.148(10) & 15.941 & 16.059 & 16.149 & 16.214 \\
      $G_0W_0$    & 14.338 & 14.466 & 14.753 & 14.872 & 23.647 & 23.626 & 23.875 & 23.988\phh & 15.679 & 15.868 & 16.237 & 16.453 \\
      qs$GW$      & 14.142 & 14.446 & 14.621 & 14.686 & 23.248 & 23.426 & 23.550 & 23.605\phh & 16.001 & 16.144 & 16.349 & 16.469 \\
      G$_0$F(2)   & 13.861 & 13.913 & 14.102 & 14.176 & 23.377 & 23.257 & 23.385 & 23.447\phh & 14.280 & 14.437 & 14.685 & 14.815 \\
      $G_0T_0$    & 14.117 & 14.117 & 14.275 &     \x & 24.107 & 24.051 & 24.163 &     \x\phh & 15.334 & 15.466 & 15.721 &     \x \\
      \hline                                                                                    
      Mol.        & \mc{4}{c}{Hydrogen fluoride (\ce{HF})} & \mc{8}{c}{Neon (\ce{Ne})} \\
      State/Conf. & \mc{4}{c}{$1~^{2}\mathrm{\Sigma}^+$/$(3\sigma)^{-1}$} & \mc{4}{c}{$1~^{2}\mathrm{P}$/$(2p)^{-1}$} & \mc{4}{c}{$1~^{2}\mathrm{S}$/$(2s)^{-1}$}  \\
      Exp.        & \mc{4}{c}{19.90 \cite{Bieri_1980}} & \mc{4}{c}{21.57\cite{Svensson_1988}} & \mc{4}{c}{48.46\cite{Svensson_1988}} \\
      \cline{2-5} \cline{6-9} \cline{10-13}
      CC2         & 18.740 & 18.814 & 18.908 & 18.982 & 19.874 & 20.017 & 20.144 & 20.236 & 47.483 & 47.265 & 47.187 & 47.207 \\
      CCSD        & 19.777 & 19.861 & 19.946 & 20.021 & 21.030 & 21.168 & 21.326 & 21.432 & 48.735 & 48.363 & 48.426 & 48.494 \\
      CC3         & 19.980 & 20.040 & 20.050 & 20.100 & 21.353 & 21.417 & 21.449 & 21.522 & 48.652 & 48.263 & 48.145 & 48.168 \\
      CCSDT       & 19.933 & 19.989 & 19.995 & 20.045 & 21.304 & 21.367 & 21.398 & 21.473 & 48.725 & 48.330 & 48.229 & 48.270 \\
      CC4         & 19.986 & 20.051 & 20.065 & 20.114 & 21.375 & 21.434 & 21.473 & 21.546 & 48.829 & 48.424 & 48.316 & 48.349 \\
      CCSDTQ      & 19.974 & 20.036 & 20.046 & 20.094 & 21.362 & 21.421 & 21.455 & 21.527 & 48.811 & 48.406 & 48.293 & 48.326 \\
      FCI         & 19.979 & 20.043 & 20.054 & 20.102 & 21.365 & 21.426 & 21.461 & 21.533 & 48.822 & 48.417 & 48.306 & 48.340 \\
      $G_0W_0$    & 19.662 & 19.812 & 20.074 & 20.259 & 20.859 & 21.104 & 21.432 & 21.655 & 47.851 & 47.785 & 47.950 & 48.085 \\
      qs$GW$      & 19.984 & 20.084 & 20.203 & 20.304 & 21.361 & 21.435 & 21.592 & 21.729 & 47.844 & 47.652 & 47.560 & 47.566 \\
      G$_0$F(2)   & 18.644 & 18.744 & 18.899 & 19.007 & 19.642 & 19.851 & 20.066 & 20.202 & 47.246 & 47.082 & 47.055 & 47.096 \\
      $G_0T_0$    & 19.312 & 19.402 & 19.551 &     \x & 20.671 & 20.847 & 21.085 &     \x & 48.966 & 48.851 & 48.886 &     \x \\
      \hline
      Mol.        & \mc{12}{c}{Water (\ce{H2O})} \\
      State/Conf. & \mc{4}{c}{$2~^{2}\mathrm{B}_1$/$(3a_1)^{-1}(1b_1)^{-1}(4a_1)^1$} & \mc{4}{c}{$2~^{2}\mathrm{A}_1$/$(1b_1)^{-2}(4a_1)^1$} & \mc{4}{c}{$3~^{2}\mathrm{B}_1$/$(3a_1)^{-1}(1b_1)^{-1}(4a_1)^1$} \\
      Exp.        &  \mc{4}{c}{27.1 \cite{Ning_2008}} &  \mc{4}{c}{27.1 \cite{Ning_2008}} &   \mc{4}{c}{} \\
      \cline{2-5} \cline{6-9} \cline{10-13}
      CC3         & 26.152 & 25.797 & 26.075 & 26.174 & 25.949 & 25.763 & 26.038 & 26.130 & 27.654 & 27.425 & 27.661 & 27.747 \\
      CCSDT       & 27.566 & 27.694 & 28.103 & 28.246 & 27.324 & 27.476 & 27.831 & 27.954 & 29.005 & 29.129 & 29.442 & 29.559 \\
      CC4         & 26.894 & 26.844 & 27.090 & 27.195 & 26.943 & 26.965 & 27.159 & 27.239 & 28.588 & 28.580 & 28.737 & 28.813 \\
      CCSDTQ      & 27.051 & 27.049 & 27.297 &     \x & 27.065 & 27.104 & 27.294 &     \x & 28.714 & 28.729 & 28.882 &     \x \\
      FCI         & 27.062 & 27.065 & 27.300 & 27.389 & 27.084 & 27.131 & 27.312 & 27.404 & 28.731 & 28.754 & 28.899 & 28.973 \\
      \hline
      Mol.        & \mc{8}{c}{Ammonia (\ce{NH3})} & \mc{4}{c}{Methane (\ce{CH4})} \\
      State/Conf. & \mc{4}{c}{$2~^{2}\mathrm{A}_1$/$(3a_1)^{-2}(4a_1)^{1}$} & \mc{4}{c}{$2~^{2}\mathrm{E}$/$(3a_1)^{-2}(3e)^{1}$} & \mc{4}{c}{$2~^{2}\mathrm{T}_2$/$(1t_2)^{-2}(3a_1)^1$} \\
      Exp.        &  &  &  &  &  &  &  &  & \mc{4}{c}{29.2\cite{Gothe_1991}} \\
      \cline{2-5} \cline{6-9} \cline{10-13}
      CC3         & 23.112 & 23.126 & 23.367 & 23.440 & 25.489 & 25.220 & 25.418 & 25.471 & 28.102 & 28.188 & 28.388 & 28.445\ph \\
      CCSDT       & 23.866 & 24.101 & 24.408 & 24.503 & 25.881 & 25.882 & 26.113 & 26.189 & 28.210 & 28.415 & 28.643 & 28.713\ph \\
      CC4         & 23.579 & 23.764 & 23.952 &     \x & 25.666 & 25.618 & 25.743 &     \x & 27.922 & 28.111 & 28.271 &     \x\ph \\
      CCSDTQ      & 23.631 & 23.818 & 24.003 &     \x & 25.688 & 25.648 & 25.773 &     \x & 27.931 & 28.123 & 28.282 &     \x\ph \\
      FCI         & 23.630 & 23.829 & 24.004 & 24.061 & 25.685 & 25.655 & 25.771 & 25.815 & 27.859 & 28.108 & 28.238 & 28.277(5) \\
      \hline
      Mol.        & \mc{8}{c}{Hydrogen fluoride (\ce{HF})} & \mc{4}{c}{Neon (\ce{Ne})} \\
      State/Conf. & \mc{4}{c}{$2~^{2}\mathrm{\Sigma}^+$/$(1\pi)^{-2}(4\sigma)^1$} & \mc{4}{c}{$1~^{2}\mathrm{\Delta}$/$(1\pi)^{-2}(4\sigma)^1$} & \mc{4}{c}{$2~^{2}\mathrm{P}$/$(2p)^{-2}(3s)^{1}$} \\
      Exp.        &  &  &  &  &  &  &  &  & \mc{4}{c}{49.16\cite{Joshi_2006}} \\
      \cline{2-5} \cline{6-9} \cline{10-13}
      CC3         & 31.076 & 30.636 & 30.916 & 31.039 & 32.872 & 32.516 & 32.749 & 32.852 & 46.502 & 46.690 & 46.436 & 46.343 \\
      CCSDT       & 32.849 & 32.917 & 33.356 & 33.531 & 34.845 & 34.885 & 35.218 & 35.365 & 49.774 & 49.917 & 50.197 & 50.322 \\
      CC4         & 32.110 & 31.981 & 32.210 &     \x & 34.309 & 34.181 & 34.304 & 34.399 & 48.932 & 48.980 & 48.920 & 48.962 \\
      CCSDTQ      & 32.312 & 32.228 & 32.466 & 32.603 & 34.503 & 34.403 & 34.528 & 34.631 & 49.258 & 49.283 & 49.313 & 49.394 \\
      FCI         & 32.347 & 32.257 & 32.474 & 32.605 & 34.547 & 34.445 & 34.554 & 34.648 & 49.339 & 49.349 & 49.343 & 49.414 \\
    \end{tabular}
  \end{ruledtabular}
\end{table*}
\end{squeezetable}

\subsection{10-electron molecules: \ce{Ne}, \ce{HF}, \ce{H2O}, \ce{NH3}, and \ce{CH4}}
\label{sec:10_elec}

\begin{figure}
  \centering
  \includegraphics[width=\linewidth]{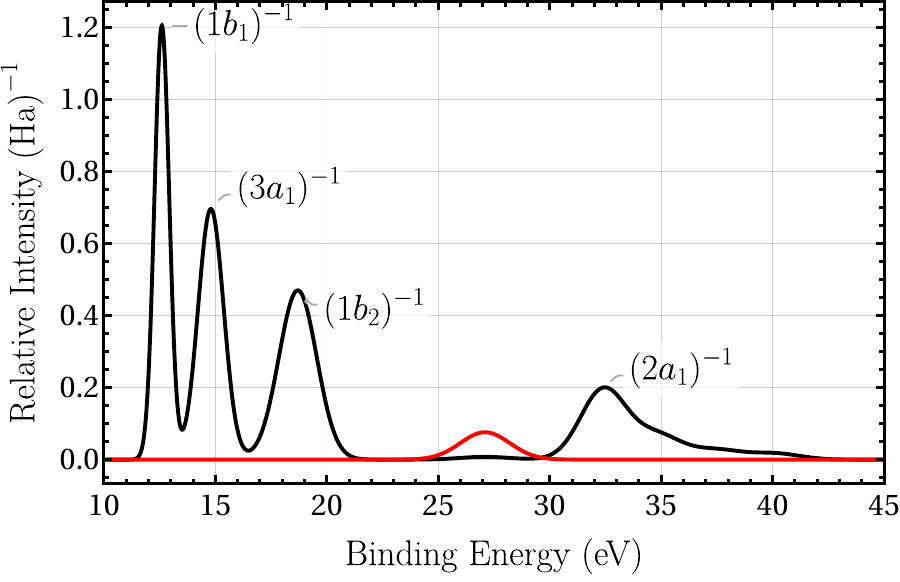}
  \caption{
  Gaussian fit of the experimental ionization spectrum of water in gas phase measured by Ning and coworkers. The fitting parameters can be found in Ref.~\onlinecite{Ning_2008}. The red peak at \SI{27.1}{\eV} has been magnified by a factor 10.}
   \label{fig:spectrum}
\end{figure}

The water molecule has been extensively studied experimentally using photoionization and electron impact spectroscopy. \cite{Cambi_1984, Banna_1986, Ning_2008}
For example, a high-resolution spectrum of liquid water is crucial as a first step for understanding the photoelectron spectra of aqueous phases. \cite{Winter_2004,Seidel_2016}
On the other hand, its gas-phase ionization spectrum is now well understood. 
The experimental ionization spectrum of water is plotted in Fig.~\ref{fig:spectrum}, serving as a representative example to illustrate the following discussion.
The first three sharp peaks at 12.6, 14.8, and \SI{18.7}{\eV} are associated with electron detachments from the three outer-valence orbitals, $1b_1$, $3a_1$, and $1b_2$, respectively. \cite{Ning_2008}
Then, a broader yet intense peak corresponding to the fourth ionization, $(2a_1)^{-1}$, is found at \SI{32.4}{\eV} surrounded by several close-lying satellite peaks.
Additionally, there is a smaller broad satellite peak at \SI{27.1}{\eV} (magnified red peak in Fig.~\ref{fig:spectrum}).

Table \ref{tab:tab1} gathers the FCI reference values corresponding to the three lowest satellites identified in our study. 
The first two, which are of $^{2}\mathrm{B}_1$ and $^{2}\mathrm{A}_1$ symmetries, lie close to each other around \SI{27.1}{\eV}.
The third satellite is of $^{2}\mathrm{B}_1$ symmetry and is found at slightly higher energy, approximately \SI{29}{\eV}.
The ordering and the absolute energies of the three satellites align well with previous SAC-CI results reported by Ning \etal \cite{Ning_2008}
In addition, they showed that, at the ADC(3) level, the energy of the $2~^{2}\mathrm{A}_1$ state is overestimated by approximately \SI{2.7}{\eV} while the $2~^{2}\mathrm{B}_1$ and $3~^{2}\mathrm{B}_1$ states are missing.
It is also worth noting that early CI and Green's function studies had qualitatively predicted the $2~^{2}\mathrm{A}_1$ satellite. \cite{Cederbaum_1974,Arneberg_1982}
Finally, we do not consider the broad peak at \SI{32.4}{\eV} here because it is technically out of reach for our current SCI implementation.
However, it has been studied by Mejuto-Zaera and coworkers who have shown that vertex corrections are required to correctly describe this complex part of the spectrum where strong many-body effects are at play. \cite{Mejuto-Zaera_2021}

For the three satellites of water, CCSDTQ is in near-perfect agreement with FCI in all basis sets with errors inferior to \SI{0.03}{\eV}.
CC4 is slightly worse than CCSDTQ but is still an excellent approximation given its lower computational cost and its approximate treatment of quadruples.
The CCSDT satellite energies are overestimated by approximately \SI{0.5}{\eV}, while CC3 appears to struggle for this system.
Indeed, the CC3 energies are badly underestimated with errors up to \SI{1.5}{\eV}, and the ordering of the first two satellites is wrongly predicted. \cite{Moitra_2022}
Finally, CCSD and CC2 are not considered for satellites as their poor performance (wrong by several \si{\eV}) makes the assignment of these states extremely challenging.

The remainder of this section is concerned with four molecules isoelectronic to water, namely \ce{CH4}, \ce{NH3}, \ce{HF}, and \ce{Ne}.
For each of these molecules, Table \ref{tab:tab1} provides FCI reference values for the IPs corresponding to the two outermost valence orbitals.
In addition, two satellite energies are reported for hydrogen fluoride and ammonia while one satellite is presented for methane and neon.
Experimental values for the IPs of these four molecules have been measured multiple times and are reported in Table \ref{tab:tab1}. \cite{Banna_1975,Banna_1975a,Bieri_1980,Campbell_1982,Bawagan_1988,Clark_1989,Clark_1990,Edvardsson_1999,Yencha_1999,Buttersack_2019}

Yencha and coworkers measured the inner-valence photoelectron spectrum of \ce{HF}.
It displays a well-defined peak around \SI{33}{\eV} which appears in close agreement with the FCI energies for the $2~^{2}\mathrm{\Sigma}^+$ state. \cite{Yencha_1999}
In addition, a doubly-degenerate satellite of $^{2}\mathrm{\Delta}$ symmetry has also been computed.
In the various \ce{NH3} ionization spectra reported in the literature, there is no satellite peak around \SI{24}{\eV} which may correspond to the $2~^{2}\mathrm{A}_1$ and $2~^{2}\mathrm{E}$ FCI states. \cite{Bawagan_1988,Edvardsson_1999,Ishida_2002}
Nevertheless, the FCI energies align well with the SAC-CI energies of Ishida and coworkers who predicted that these two satellite states have very low intensity. \cite{Ishida_2002}
The first satellite observed in the inner-valence region of the photoionization spectrum of \ce{CH4} is a very weak and broad peak at \SI{29.2}{\eV}. \cite{Gothe_1991}
This peak is also measured at \SI{28.56}{\eV} using electron momentum spectroscopy experiments. \cite{Clark_1990}
The energy of the first satellite calculated at the FCI level, and associated with the $(2t_2)^{-2}(3a_1)^1$ process, compares well with the experimental data.
Finally, the lowest-energy satellite state of neon is also reported along with the corresponding experimental value measured by Joshi and co-workers. \cite{Joshi_2006}
It is worth noting that CC3 behaves similarly in \ce{Ne}, \ce{HF} and \ce{H2O}, yet it appears to be a much better approximation for the satellite states of \ce{NH3} and \ce{CH4}.

Among the 12 IPs computed for this series of molecules, 11 of them have a weight larger than 0.85 on the 1h dominant configuration (in the aug-cc-pVDZ basis set).
Only the $3~^{2}\mathrm{A}_1$ state of ammonia has a quite smaller weight (0.58) on the corresponding 1h determinant.
This exemplifies the breakdown of the orbital picture in the inner-valence ionization spectrum, \cite{Cederbaum_1974,Schirmer_1977,Cederbaum_1977,Schirmer_1978,Schirmer_1978a,Schirmer_1978b,Domcke_1978,Cederbaum_1978,Cederbaum_1980,vonNiessen_1980,vonNiessen_1981,Walter_1981,Schirmer_1983a,Cederbaum_1986} which signature is a significant weight on both 1h and 2h1p configurations, hence preventing us from assigning the solution as a clear IP or satellite.
The performance of the various approximations for IPs will be statistically gauged in Sec.~\ref{sec:stat}.

\begin{table}
  \caption{Satellite transition energies of ammonia and water computed with Green's function methods in the aug-cc-pVDZ basis set.
  The FCI values are reported for comparison purposes.}
   \label{tab:tab2}
  \begin{ruledtabular}
    \begin{tabular}{llcccc}
      Molecule & State                &  Method   & Diag. element & Eigenvalue &   FCI  \\
      \hline
               &                      & G$_0$F(2) &         24.324               &   24.328   &        \\
      \ce{NH3} & $2~^{2}\mathrm{A}_1$ & $G_0W_0$  &         24.408               &   24.410   & 23.829 \\
               &                      & $G_0T_0$  &         40.441               &   40.444   &        \\
      \hline
               &                      & G$_0$F(2) &         24.977               &   24.977   &        \\
      \ce{NH3} & $2~^{2}\mathrm{E}$   & $G_0W_0$  &         24.997               &   24.997   & 25.655 \\
               &                      & $G_0T_0$  &         41.094               &   41.094   &        \\
      \hline
               &                      & G$_0$F(2) &         30.759               &   30.759   &        \\
      \ce{H2O} & $2~^{2}\mathrm{B}_1$ & $G_0W_0$  &         30.846               &   30.846   & 27.065 \\
               &                      & $G_0T_0$  &             \x               &       \x   &        \\
      \hline
               &                      & G$_0$F(2) &         28.683               &   28.683   &        \\
      \ce{H2O} & $2~^{2}\mathrm{A}_1$ & $G_0W_0$  &         28.770               &   28.770   & 27.131 \\
               &                      & $G_0T_0$  &             \x               &       \x   &        \\
      \hline
               &                      & G$_0$F(2) &         30.759               &   30.781   &        \\
      \ce{H2O} & $3~^{2}\mathrm{B}_1$ & $G_0W_0$  &         30.863               &   30.867   & 28.754 \\
               &                      & $G_0T_0$  &             \x               &       \x   &        \\
    \end{tabular}
  \end{ruledtabular}
\end{table}

\subsection{Satellite in Green's functions methods}

Thus far, we have exclusively assessed the performance of different rungs of the CC hierarchy. 
Although shake-up transition energies can also be computed within the Green's function framework, the task is notably more challenging, especially when compared to the more straightforward nature of IP-EOM-CC. 
This complexity arises from the fact that satellites, existing as non-linear solutions of the quasiparticle equation, \cite{Martin_2016} prove much more difficult to converge using Newton-Raphson algorithms than the quasiparticle solutions, which are representative of typical IPs.
Fortunately, an alternative and equivalent pathway exists where one solves a larger linear eigenvalue problem instead of solving the non-linear quasiparticle equation. \cite{Bintrim_2021,Schirmer_2018,Backhouse_2020b,Monino_2021,Monino_2022,Tolle_2023,Monino_2023,Tolle_2023a,Scott_2023}
In such a case, satellites are obtained as higher-energy roots via diagonalization of the so-called ``upfolded'' matrix built in the basis of the 2h1p and two-particle-one-hole (2p1h) configurations in addition to the 1h and 1p configurations.

The satellite energies of \ce{H2O} and \ce{NH3} computed with $G_0W_0$, G$_0$F(2), and $G_0T_0$ are presented in Table \ref{tab:tab2}.
qs$GW$ is not considered here as its static approximation naturally discards all the satellite solutions.
The third column shows the diagonal elements of the upfolded matrix associated with the 2h1p configurations, while the fourth column displays the associated eigenvalues.
One can immediately observe that the eigenvalues do not improve upon the diagonal elements.
This is due to the lack of higher-order (such as 3h2p) configurations that are essential to correlate satellites.
This parallels the description of double excitations which require at least triple excitations (\ie~3h3p configurations) in addition to the 2h2p configurations to correlate doubly-excited states. \cite{Starcke_2006,Loos_2019,doCasal_2023}
Regarding the two satellites of ammonia, the $T$-matrix zeroth-order elements are utterly inaccurate.
This discrepancy arises because, at the $T$-matrix level, satellite energies are described as the sum of a Koopmans' electron attachment energy (1p configuration) and a double electron detachment energy [two-hole (2h) configuration] stemming from the particle-particle random-phase approximation. \cite{Schuck_Book,Scuseria_2013,Peng_2013,Berkelbach_2018,Monino_2023}
On the other hand, $G_0W_0$ and G$_0$F(2) offer decent estimates of these satellite energies.
The difference in performance of $G_0T_0$ and $G_0W_0$ can be understood in terms of the scattering channels that are accounted for in each of these approximations. We refer the interested reader to Ref.~\onlinecite{Orlando_2023b} and references therein.

The remaining three rows of Table \ref{tab:tab2} contain the energies corresponding to the three satellites of water discussed previously.
The $2~^{2}\mathrm{A}_1$ state is the easiest to identify as its 2h1p dominant configuration clearly corresponds to the $(1b_1)^{-2}(4a_1)^1$ process.
The eigenvectors associated with the $2~^{2}\mathrm{B}_1$ and $3~^{2}\mathrm{B}_1$ states, which correspond to the $(3a_1)^{-1}(1b_1)^{-1}(4a_1)^1$ and $(1b_1)^{-1}(3a_1)^{-1}(4a_1)^1$ processes respectively, are nearly degenerate and highly entangled. This is thus harder, if not impossible, to assign these states.

Because of these assignment problems, the $G_0W_0$, G$_0$F(2), and $G_0T_0$ satellite energies have not been computed for the other molecules considered in this study.
To alleviate this issue, there is a notable appeal for a self-energy approximation including vertex corrections capable of effectively addressing satellite states.
As mentioned previously, Green's-function-based methods such as the 2ph-TDA \cite{Schirmer_1978} and ADC(3) \cite{Schirmer_1983a,Trofimov_2005,Banerjee_2019} (first named extended 2ph-TDA \cite{Walter_1981}) have shown success in qualitatively modeling the inner-valence region of experimental spectra. \cite{Cederbaum_1974,Schirmer_1977,Cederbaum_1977,Schirmer_1978,Schirmer_1978a,Schirmer_1978b,Domcke_1978,Cederbaum_1978,Cederbaum_1980,vonNiessen_1980,vonNiessen_1981,Walter_1981,Schirmer_1983,Cederbaum_1986}
Sokolov's recent multi-reference ADC(2) scheme \cite{Sokolov_2018} is also a promising avenue. 
In particular, it has shown potential in describing the satellites of the carbon dimer (see below). \cite{Chatterjee_2019,Chatterjee_2020}
While a detailed quantitative analysis of these approaches on the present benchmark set would be interesting, it lies beyond the scope of this study.

\begin{squeezetable}
\begin{table*}
  \caption{Valence ionizations and satellite transition energies (in \si{\eV}) of the 14-electron series for various methods and basis sets. 
  AVXZ stands for aug-cc-pVXZ (where X = D, T, and Q). Selected experimental values are also reported.}
  \label{tab:tab3}
  \begin{ruledtabular}
    \begin{tabular}{rccccccccccccc}
      & \mc{4}{c}{Basis} & \mc{4}{c}{Basis} & \mc{4}{c}{Basis} \\
      \cline{2-5} \cline{6-9} \cline{10-13}
      Methods & \mcc{6-31$+$G$^{*}$} & \mcc{AVDZ} & \mcc{AVTZ} & \mcc{AVQZ} & \mcc{6-31$+$G$^{*}$} & \mcc{AVDZ} & \mcc{AVTZ} & \mcc{AVQZ} & \mcc{6-31$+$G$^{*}$} & \mcc{AVDZ} & \mcc{AVTZ} & \mcc{AVQZ}  \\
      \hline
      Mol.        & \mc{12}{c}{Boron fluoride (\ce{BF})} \\
      State/Conf. & \mc{4}{c}{$1~^{2}\Sigma^+$/$(5\sigma)^{-1}$} & \mc{4}{c}{$2~^{2}\Pi$/$(1\pi)^{-1}$} & \mc{4}{c}{$2~^{2}\Sigma^+$/$(4\sigma)^{-1}$} \\
      Exp.        & \mc{4}{c}{11.06 \cite{Hildenbrand_1971}} &  &  &  &  &  &  &  &  \\
      \cline{2-5} \cline{6-9} \cline{10-13}
      CC2         & 10.751 & 10.824 & 10.944 & 10.987 & 17.136 & 17.274 & 17.385 & 17.467 & 19.767 & 19.957 & 19.962 & 20.041\ph \\
      CCSD        & 11.080 & 11.154 & 11.250 & 11.279 & 17.920 & 18.028 & 18.172 & 18.246 & 21.017 & 21.208 & 21.253 & 21.342\ph \\
      CC3         & 10.914 & 11.004 & 11.100 & 11.127 & 18.606 & 18.722 & 18.743 & 18.794 & 20.745 & 20.910 & 20.918 & 20.984\ph \\
      CCSDT       & 10.969 & 11.057 & 11.157 & 11.185 & 18.474 & 18.584 & 18.622 & 18.677 & 20.708 & 20.866 & 20.875 & 20.946\ph \\
      CC4         & 10.965 & 11.053 & 11.148 & 11.175 & 18.475 & 18.593 & 18.626 & 18.679 & 20.770 & 20.920 & 20.917 & 20.981\ph \\
      CCSDTQ      & 10.967 & 11.054 & 11.150 &     \x & 18.453 & 18.569 & 18.599 &     \x & 20.734 & 20.882 & 20.874 &     \x\ph \\
      FCI         & 10.966 & 11.054 & 11.149 & 11.175 & 18.466 & 18.581 & 18.612 & 18.664 & 20.765 & 20.913 & 20.906 & 20.970(1) \\
      $G_0W_0$    & 11.053 & 11.117 & 11.325 & 11.420 & 18.237 & 18.456 & 18.743 & 18.949 & 21.142 & 21.402 & 21.567 & 21.778\ph \\
      qs$GW$      & 10.862 & 10.989 & 11.167 & 11.240 & 18.513 & 18.662 & 18.784 & 18.899 & 21.389 & 21.583 & 21.585 & 21.701\ph \\
      G$_0$F(2)   & 10.859 & 10.915 & 11.052 & 11.114 & 16.958 & 17.132 & 17.328 & 17.454 & 19.654 & 19.878 & 19.955 & 20.079\ph \\
      $G_0T_0$    & 10.821 & 10.856 & 10.955 &     \x & 17.947 & 18.105 & 18.304 &     \x & 20.739 & 20.955 & 21.041 &     \x\ph \\
      \hline
      Mol.        & \mc{12}{c}{Carbon monoxide (\ce{CO})} \\
      State/Conf. & \mc{4}{c}{$1~^{2}\Sigma^+$/$(5\sigma)^{-1}$} & \mc{4}{c}{$1~^{2}\Pi$/$(1\pi)^{-1}$} & \mc{4}{c}{$2~^{2}\Sigma^+$/$(4\sigma)^{-1}$} \\
      Exp.        & \mc{4}{c}{14.01\cite{Svensson_1991}} & \mc{4}{c}{17.0\cite{Svensson_1991}} & \mc{4}{c}{19.7\cite{Svensson_1991}} \\
      \cline{2-5} \cline{6-9} \cline{10-13}
      CC2         & 13.550 & 13.584 & 13.748 & 13.809 & 16.289 & 16.349 & 16.505 & 16.581\ph & 18.175 & 18.316 & 18.400 & 18.464\ph \\
      CCSD        & 13.948 & 13.998 & 14.190 & 14.246 & 16.793 & 16.865 & 17.024 & 17.095\ph & 19.501 & 19.657 & 19.790 & 19.867\ph \\
      CC3         & 13.614 & 13.697 & 13.863 & 13.912 & 16.826 & 16.902 & 17.018 & 17.075\ph & 19.512 & 19.664 & 19.744 & 19.807\ph \\
      CCSDT       & 13.693 & 13.770 & 13.952 & 14.005 & 16.762 & 16.838 & 16.960 & 17.016\ph & 19.347 & 19.498 & 19.583 & 19.647\ph \\
      CC4         & 13.678 & 13.760 & 13.933 & 13.984 & 16.751 & 16.835 & 16.955 & 17.009\ph & 19.410 & 19.566 & 19.653 & 19.715\ph \\
      CCSDTQ      & 13.679 & 13.761 & 13.935 &     \x & 16.755 & 16.837 & 16.958 &     \x\ph & 19.376 & 19.532 & 19.616 &     \x\ph \\
      FCI         & 13.670 & 13.752 & 13.925 & 13.975 & 16.762 & 16.845 & 16.966 & 17.017(2) & 19.393 & 19.550 & 19.637 & 19.699(1) \\
      $G_0W_0$    & 14.461 & 14.467 & 14.777 & 14.915 & 16.677 & 16.762 & 17.083 & 17.264\ph & 19.869 & 20.045 & 20.300 & 20.485\ph \\
      qs$GW$      & 13.980 & 14.080 & 14.318 & 14.416 & 16.836 & 16.932 & 17.124 & 17.231\ph & 19.899 & 20.071 & 20.191 & 20.298\ph \\
      G$_0$F(2)   & 13.856 & 13.857 & 14.067 & 14.154 & 16.134 & 16.204 & 16.422 & 16.534\ph & 18.165 & 18.317 & 18.460 & 18.564\ph \\
      $G_0T_0$    & 14.163 & 14.143 & 14.324 &     \x & 16.422 & 16.470 & 16.666 &     \x\ph & 19.333 & 19.481 & 19.613 &     \x\ph \\
      \hline
      Mol.        & \mc{12}{c}{Dinitrogen (\ce{N2})} \\
      State/Conf. & \mc{4}{c}{$1~^{2}\Sigma_g^+$/$(3\sigma_g)^{-1}$} & \mc{4}{c}{$1~^{2}\Pi_u$/$(1\pi_u)^{-1}$} & \mc{4}{c}{$1~^{2}\Sigma_u^+$/$(2\sigma_u)^{-1}$} \\
      Exp.        & \mc{4}{c}{15.580\cite{Baltzer_1992}} & \mc{4}{c}{16.926\cite{Baltzer_1992}} & \mc{4}{c}{18.751\cite{Baltzer_1992}} \\
      \cline{2-5} \cline{6-9} \cline{10-13}
      CC2         & 14.613 & 14.649 & 14.814 & 14.877 & 16.932 & 16.943 & 17.104 & 17.178 & 17.803 & 17.862 & 17.991 & 18.037 \\
      CCSD        & 15.382 & 15.424 & 15.641 & 15.709 & 17.065 & 17.087 & 17.228 & 17.287 & 18.654 & 18.721 & 18.931 & 18.991 \\
      CC3         & 15.282 & 15.349 & 15.519 & 15.574 & 16.669 & 16.719 & 16.837 & 16.885 & 18.598 & 18.680 & 18.849 & 18.899 \\
      CCSDT       & 15.270 & 15.333 & 15.517 & 15.574 & 16.765 & 16.812 & 16.950 & 17.001 & 18.502 & 18.585 & 18.763 & 18.816 \\
      CC4         & 15.220 & 15.293 & 15.471 & 15.526 & 16.770 & 16.821 & 16.940 & 16.987 & 18.403 & 18.493 & 18.669 & 18.720 \\
      CCSDTQ      & 15.237 & 15.309 & 15.487 &     \x & 16.764 & 16.815 & 16.936 &     \x & 18.429 & 18.519 & 18.696 &     \x \\
      FCI         & 15.235 & 15.308 & 15.486 & 15.541 & 16.759 & 16.811 & 16.933 & 16.981 & 18.427 & 18.516 & 18.692 & 18.742 \\
      $G_0W_0$    & 15.959 & 15.984 & 16.350 & 16.519 & 16.781 & 16.790 & 17.093 & 17.259 & 19.515 & 19.558 & 19.862 & 20.000 \\
      qs$GW$      & 15.575 & 15.663 & 15.914 & 16.020 & 16.640 & 16.706 & 16.903 & 17.006 & 19.125 & 19.221 & 19.425 & 19.513 \\
      G$_0$F(2)   & 14.824 & 14.845 & 15.080 & 15.181 & 16.956 & 16.952 & 17.158 & 17.261 & 17.974 & 18.020 & 18.201 & 18.274 \\
      $G_0T_0$    & 15.494 & 15.502 & 15.722 &     \x & 16.673 & 16.653 & 16.820 &     \x & 18.993 & 19.021 & 19.190 &     \x \\
      \hline
      Mol.        & \mc{4}{c}{Boron Fluoride (\ce{BF})} &  \mc{8}{c}{} \\
      State/Conf. & \mc{4}{c}{$1~^{2}\Pi$/$(5\sigma)^{-2}(2\pi)^1$} &  \mc{8}{c}{} \\
      Exp.        & \mc{4}{c}{} &  \mc{8}{c}{} \\
      \cline{2-5} 
      CC3         & 17.494 & 17.541 & 17.607 & 17.637 &        &        &        &        &        &        &        &        \\
      CCSDT       & 17.410 & 17.462 & 17.532 & 17.567 &        &        &        &        &        &        &        &        \\
      CC4         & 17.293 & 17.345 & 17.393 & 17.419 &        &        &        &        &        &        &        &        \\
      CCSDTQ      & 17.303 & 17.355 & 17.405 &     \x &        &        &        &        &        &        &        &        \\
      FCI         & 17.297 & 17.346 & 17.392 & 17.417 &        &        &        &        &        &        &        &        \\
      \hline
      Mol.        & \mc{12}{c}{Carbon monoxide (\ce{CO})} \\
      State/Conf. & \mc{4}{c}{$2~^{2}\Pi$/$(5\sigma)^{-2}(2\pi)^{1}$} & \mc{4}{c}{$2~^{2}\Sigma^+$/$(1\pi)^{-1}(5\sigma)^{-1}(2\pi)^{1}$} & \mc{4}{c}{$1~^{2}\Delta$/$(1\pi)^{-1}(5\sigma)^{-1}(2\pi)^{1}$} \\
      Exp.        & \mc{4}{c}{22.7\cite{Svensson_1991}} & \mc{4}{c}{23.7\cite{Svensson_1991}} & \mc{4}{c}{}   \\
      \cline{2-5} \cline{6-9} \cline{10-13}
      CC3         & 23.406 & 23.507 & 23.597\ph & 23.640\ph & 23.640\ph & 23.729\ph & 23.839\ph & 23.881\ph & 23.730 & 23.814 & 23.926 & 23.968\ph \\
      CCSDT       & 23.205 & 23.313 & 23.441\ph & 23.507\ph & 23.381\ph & 23.472\ph & 23.602\ph & 23.669\ph & 23.417 & 23.503 & 23.647 & 23.713\ph \\
      CC4         & 22.862 & 22.957 & 22.997\ph & 23.040\ph & 23.102\ph & 23.166\ph & 23.193\ph & 23.236\ph & 23.143 & 23.206 & 23.251 & 23.293x\ph \\
      CCSDTQ      & 22.841 & 22.937 & 22.995\ph &     \x\ph & 23.101\ph & 23.167\ph & 23.209\ph &     \x\ph & 23.141 & 23.205 & 23.264 &     \x\ph \\
      FCI         & 22.791 & 22.889 & 22.908(1) & 22.962(3) & 23.074(1) & 23.140(2) & 23.194(1) & 23.232(1) & 23.114 & 23.181 & 23.233 & 23.271(1) \\
      \hline
      Mol.        & \mc{12}{c}{Dinitrogen (\ce{N2})}  \\
      State/Conf. & \mc{4}{c}{$1~^{2}\Pi_g$/$(3\sigma_g)^{-2}(4\pi_g)^1$}& \mc{4}{c}{$1~^{2}\Sigma_u^+$/$(3\sigma_g)^{-1}(3\pi_u)^{-1}(4\pi_g)^1$} & \mc{4}{c}{$1~^{2}\Sigma_u^-$/$(3\sigma_g)^{-1}(3\pi_u)^{-1}(4\pi_g)^1$} \\
      Exp.        & \mc{4}{c}{24.788 \cite{Baltzer_1992}} & \mc{4}{c}{25.514 \cite{Baltzer_1992}} & \mc{4}{c}{} \\
      \cline{2-5} \cline{6-9} \cline{10-13}
      CC3         & 25.280 & 25.331 & 25.495\ph & 25.535\ph & 25.656 & 25.699 & 25.856 & 25.908 & 26.584 & 26.599 & 26.686 & 26.723 \\
      CCSDT       & 24.945 & 25.008 & 25.232\ph & 25.304\ph & 25.405 & 25.453 & 25.643 & 25.721 & 26.209 & 26.250 & 26.362 & 26.427 \\
      CC4         & 24.394 & 24.458 & 24.575\ph & 24.621\ph & 25.099 & 25.142 & 25.235 & 25.288 & 25.990 & 26.012 & 26.022 & 26.058 \\
      CCSDTQ      & 24.363 & 24.431 & 24.574\ph &     \x\ph & 25.088 & 25.134 & 25.238 &     \x & 25.721 & 25.756 & 25.762 &     \x \\
      FCI         & 24.277 & 24.348 & 24.470(1) & 24.519(1) & 25.054 & 25.103 & 25.199 &     \x & 25.658 & 25.695 & 25.689 &     \x \\
    \end{tabular}
  \end{ruledtabular}
\end{table*}
\end{squeezetable}

\subsection{14-electron molecules: \ce{N2}, \ce{CO}, and \ce{BF}}

The nitrogen and carbon monoxide molecules have been extensively studied both experimentally \cite{Potts_1974,Asbrink_1974a,Asbrink_1974b,Banna_1976,Norton_1978,Svensson_1991,Baltzer_1992} and theoretically. \cite{Schirmer_1977,Bagus_1977,Kosugi_1979,Honjou_1981,Schirmer_1983a,Langhoff_1988,Morrison_1992,Ehara_1998,Kamiya_2006,Dutta_2015}
Their ionization spectra are similar as they exhibit three sharp and intense peaks, corresponding to Koopmans' states, below \SI{20}{\eV}.
Their respective fourth IP, corresponding to electron detachment from the orbital $2\sigma_g$ for \ce{N2} and $3\sigma$ for \ce{CO}, lies above \SI{30}{\eV}. 
Several peaks can be found below these ionizations, \ie~between 20 and \SI{30}{\eV}. \cite{Asbrink_1974a,Asbrink_1974b,Svensson_1991}
These correspond to satellite states associated with the three outer-valence orbitals.
Note that Schirmer \etal~have shown (using the 2ph-TDA method \cite{Schirmer_1978}) that the quasiparticle approximation breaks down in the region of the fourth ionizations of \ce{CO} and \ce{N2}. \cite{Schirmer_1977,Schirmer_1983a}
However, as shown below, the peaks between 20 and \SI{30}{\eV} have a well-defined satellite character.

Baltzer \etal~produced, using \ce{He(II)} photoelectron spectroscopy, accurate experimental values for the outer-valence IPs (see Table~\ref{tab:tab3}) and the first satellite peaks of \ce{N2}. \cite{Baltzer_1992}
In particular, they reported a value of \SI{25.514}{\eV} for an intense satellite peak, as well as \SI{24.788}{\eV} for a very weak peak.
These peaks were assigned $^{2}\Sigma_u^+$ and $^{2}\Pi_g$ symmetry, respectively, based on CI calculations.
Note that the $^{2}\Pi_g$ satellite peak is more intense when measured by resonance Auger spectroscopy. \cite{Svensson_1991}
We report FCI values for both satellites as well as a slightly higher third one with $^{2}\Sigma_u^-$ symmetry.
This latter state is not observed experimentally but plays an important role nonetheless as it is involved in the dissociation pathways between the $^{2}\Sigma_u^+$ and $^{4}\Pi_u$ states. \cite{Langhoff_1988}

Similar to its isoelectronic \ce{N2} molecule, \ce{CO} exhibits shake-up peaks between the $(4\sigma)^{-1}$ and $(3\sigma)^{-1}$ ionizations.
Using monochromatized X-ray excited photoelectron spectroscopy, Svensson \etal~observed an intense $^{2}\Sigma^+$ satellite peak with energy \SI{23.7}{\eV} as well as a weak $^2\Pi$ satellite at \SI{22.7}{\eV}. \cite{Svensson_1991}
This is in agreement with older \ce{He(II)} photoelectron spectroscopy experiments done by Asbrink and coworkers. \cite{Asbrink_1974b}
FCI values for both satellites, as well as for the higher-energy $1~^{2}\Delta$ state, are reported in Table \ref{tab:tab2}.

The performance of CC schemes for these six satellites is similar to what we have observed for the 10-electron series.
Yet, it is interesting to note that CCSDTQ seems to struggle slightly more with the $2~^{2}\Pi$ satellite of CO and the $1~^{2}\Pi_g$ and $1~^{2}\Sigma_u^-$ states of \ce{N2}.

The boron fluoride molecule is isoelectronic to \ce{CO} and \ce{N2} but its ionization spectrum is much harder to obtain experimentally because \ce{BF} is a quite non-volatile compound, meaning that the measurements have to be done at high temperatures. \cite{Hildenbrand_1971}
Yet, Hildenbrand and coworkers managed to measure its principal IP using electron impact spectroscopic and they reported a value of \SI{11.06}{\eV}.
The $1~^{2}\Sigma^+$ FCI state is in good agreement with this value.
Table \ref{tab:tab2} also displays two additional IPs and one satellite.
The order of the $^{2}\Pi$ states in \ce{BF} is reversed with respect to its isoelectronic species: the $(5\sigma)^{-2}(2\pi)^1$ satellite state has a lower energy than the $(1\pi)^{-1}$ ionization.
In this case, CCSDTQ accurately describes the satellite of $\Pi$ symmetry. 

\begin{squeezetable}
\begin{table*}
  \caption{Valence ionizations and satellite transition energies (in \si{\eV}) of the 12-electron series for various methods and basis sets. 
  AVXZ stands for aug-cc-pVXZ (where X = D, T, and Q). Selected experimental values are also reported.}
  \label{tab:tab4}
  \begin{ruledtabular}
    \begin{tabular}{rccccccccccccc}
      & \mc{4}{c}{Basis} & \mc{4}{c}{Basis} & \mc{4}{c}{Basis} \\
      \cline{2-5} \cline{6-9} \cline{10-13}
      Methods & \mcc{6-31$+$G$^{*}$} & \mcc{AVDZ} & \mcc{AVTZ} & \mcc{AVQZ} & \mcc{6-31$+$G$^{*}$} & \mcc{AVDZ} & \mcc{AVTZ} & \mcc{AVQZ} & \mcc{6-31$+$G$^{*}$} & \mcc{AVDZ} & \mcc{AVTZ} & \mcc{AVQZ}  \\
      \hline
      Mol.        & \mc{8}{c}{Lithium fluoride (\ce{LiF})} & \mc{4}{c}{Beryllium oxide (\ce{BeO})} \\
      State/Conf. & \mc{4}{c}{$1~^{2}\Pi$/$(1\pi)^{-1}$} & \mc{4}{c}{$1~^{2}\Sigma^+$/$(4\sigma)^{-1}$} & \mc{4}{c}{$1~^{2}\Pi$/$(1\pi)^{-1}$} \\
      Exp.        & \mc{4}{c}{11.50,11.67 \cite{Berkowitz_1979}} & \mc{4}{c}{11.94 \cite{Berkowitz_1979}} &  &  &  &  \\
      \cline{2-5} \cline{6-9} \cline{10-13}
      CC2         & \z9.481 & \z9.588 & \z9.804 & \z9.895 & \z9.801 & \z9.923 & 10.109 & 10.209\ph & 9.615 & \z9.712 & \z9.818 & \z9.894 \\
      CCSD        &  11.078 &  11.193 &  11.398 &  11.493 &  11.566 &  11.701 & 11.874 & 11.979\ph & 9.708 & \z9.808 & \z9.875 & \z9.941 \\
      CC3         &  11.142 &  11.222 &  11.375 &  11.449 &  11.588 &  11.682 & 11.802 & 11.883\ph & 9.976 &  10.096 &  10.243 &  10.320 \\
      CCSDT       &  11.165 &  11.247 &  11.379 &  11.452 &  11.641 &  11.738 & 11.834 & 11.914\ph & 9.750 & \z9.849 & \z9.864 & \z9.916 \\
      CC4         &  11.270 &  11.351 &  11.496 &  11.567 &  11.764 &  11.863 & 11.971 & 12.049\ph & 9.749 & \z9.850 & \z9.867 & \z9.922 \\
      CCSDTQ      &  11.234 &  11.315 &  11.453 &      \x &  11.719 &  11.816 & 11.917 &     \x\ph & 9.831 & \z9.930 & \z9.939 &      \x \\
      FCI         &  11.246 &  11.328 &  11.468 &  11.538 &  11.735 &  11.833 & 11.933 & 12.018(2) & 9.863 & \z9.962 & \z9.972 &  10.018 \\
      $G_0W_0$    &  10.797 &  10.979 &  11.384 &  11.594 &  11.339 &  11.549 & 11.915 & 12.139\ph & 9.356 & \z9.489 & \z9.727 & \z9.907 \\
      qs$GW$      &  11.249 &  11.330 &  11.575 &  11.699 &  11.809 &  11.926 & 12.119 & 12.255\ph & 9.991 &  10.079 &  10.168 &  10.251 \\
      G$_0$F(2)   & \z9.294 & \z9.445 & \z9.729 & \z9.857 & \z9.644 & \z9.812 & 10.063 & 10.200\ph & 7.957 & \z8.072 & \z8.195 & \z8.302 \\
      $G_0T_0$    &  10.512 &  10.644 &  10.923 &      \x &  10.968 &  11.120 & 11.366 &     \x\ph & 8.885 & \z8.975 & \z9.108 &      \x \\
      \hline
      Mol.        & \mc{4}{c}{Beryllium oxide (\ce{BeO})} & \mc{8}{c}{Boron nitride (\ce{BN})} \\
      State/Conf. & \mc{4}{c}{$1~^{2}\Sigma^+$/$(4\sigma)^{-1}$} & \mc{4}{c}{$1~^{2}\Pi$/$(1\pi)^{-1}$} & \mc{4}{c}{$1~^{2}\Sigma^+$/$(4\sigma)^{-1}$} \\
      Exp.        &  &  &  &  &  &  &  &  &  &  &  &  \\
      \cline{2-5} \cline{6-9} \cline{10-13}
      CC2         &  10.523 &  10.620 &  10.667 &  10.735\ph & 10.734 & 10.792 & 10.927 & 10.991 & 12.842 & 12.870 & 12.979\phh & 13.018\phh \\
      CCSD        &  10.861 &  10.987 &  11.006 &  11.082\ph & 11.776 & 11.850 & 11.971 & 12.018 & 13.571 & 13.624 & 13.697\phh & 13.720\phh \\
      CC3         &  11.128 &  11.269 &  11.370 &  11.454\ph & 11.825 & 11.941 & 12.057 & 12.101 & 13.641 & 13.718 & 13.806\phh & 13.828\phh \\
      CCSDT       &  10.830 &  10.962 &  10.916 &  10.975\ph & 11.778 & 11.871 & 11.980 & 12.019 & 13.642 & 13.716 & 13.790\phh & 13.808\phh \\
      CC4         &  10.825 &  10.959 &  10.915 &  10.977\ph & 11.681 & 11.797 & 11.902 & 11.940 & 13.534 & 13.626 & 13.700\phh & 13.718\phh \\
      CCSDTQ      &  10.923 &  11.056 &  11.007 &      \x\ph & 11.754 & 11.860 & 11.966 &     \x & 13.580 & 13.667 & 13.745\phh &     \x\phh \\
      FCI         &  10.970 &  11.103 &  11.056 &  11.115(2) & 11.767 & 11.875 & 11.980 & 12.019 & 13.571 & 13.660 & 13.729(11) & 13.710(70)  \\
      $G_0W_0$    &  10.628 &  10.798 &  10.996 &  11.200\ph & 11.423 & 11.447 & 11.752 & 11.907 & 13.154 & 13.171 & 13.447\phh & 13.590\phh \\
      qs$GW$      &  11.083 &  11.199 &  11.241 &  11.341\ph & 11.597 & 11.711 & 11.898 & 11.987 & 13.376 & 13.490 & 13.621\phh & 13.693\phh \\
      G$_0$F(2)   & \z8.499 & \z8.648 & \z8.720 & \z8.837\ph & 10.817 & 10.857 & 11.031 & 11.122 & 12.207 & 12.195 & 12.382\phh & 12.454\phh \\
      $G_0T_0$    & \z9.930 &  10.064 &  10.155 &      \x\ph & 10.988 & 10.996 & 11.159 &     \x & 12.499 & 12.512 & 12.657\phh &     \x\phh \\
      \hline
      Mol.        & \mc{4}{c}{Carbon dimer (\ce{C2})} & \mc{8}{c}{} \\
      State/Conf. & \mc{4}{c}{$1~^{2}\Pi_u$/$(2\pi_u)^{-1}$} & \mc{4}{c}{} & \mc{4}{c}{} \\
      Exp.        &  &  &  &  &  &  &  &  &  &  &  &  \\
      \cline{2-5} \cline{6-9} \cline{10-13}
      CC2         & 12.742 & 12.779 & 12.951 & 13.023 &  &  &  &  &  &  &  &  \\
      CCSD        & 12.770 & 12.830 & 12.978 & 13.030 &  &  &  &  &  &  &  &  \\
      CC3         & 11.930 & 12.058 & 12.177 & 12.215 &  &  &  &  &  &  &  &  \\
      CCSDT       & 12.289 & 12.391 & 12.540 & 12.585 &  &  &  &  &  &  &  &  \\
      CC4         & 12.231 & 12.347 & 12.472 & 12.511 &  &  &  &  &  &  &  &  \\
      CCSDTQ      & 12.225 & 12.340 & 12.471 &     \x &  &  &  &  &  &  &  &  \\
      FCI         & 12.205 & 12.323 & 12.463 & 12.497 &  &  &  &  &  &  &  &  \\
      $G_0W_0$    & 12.621 & 12.613 & 12.928 & 13.082 &  &  &  &  &  &  &  &  \\
      qs$GW$      & 12.202 & 12.344 & 12.561 & 12.656 &  &  &  &  &  &  &  &  \\
      G$_0$F(2)   & 12.882 & 12.870 & 13.078 & 13.175 &  &  &  &  &  &  &  &  \\
      $G_0T_0$    & 12.482 & 12.454 & 12.625 &     \x &  &  &  &  &  &  &  &  \\
      \hline
      Mol.        & \mc{12}{c}{Carbon dimer (\ce{C2})} \\
      State/Conf. & \mc{4}{c}{$1~^{2}\Delta_g$/$(2\pi_u)^{-2}(3\sigma_g)^1$} & \mc{4}{c}{$1~^{2}\Sigma_g^-$/$(2\pi_u)^{-2}(3\sigma_g)^1$} & \mc{4}{c}{$1~^{2}\Sigma_g^+$/$(2\pi_u)^{-2}(3\sigma_g)^1$} \\
      Exp.        &  &  &  &  &  &  &  &  &  &  &  &  \\
      \cline{2-5} \cline{6-9} \cline{10-13}
      CC3         & 14.644 & 14.713 & 14.815 & 14.850 & 14.846 & 14.957 & 15.087 & 15.123 & 15.360 & 15.353 & 15.435 & 15.460\ph \\
      CCSDT       & 14.494 & 14.568 & 14.680 & 14.729 & 14.721 & 14.833 & 14.998 & 15.051 & 15.086 & 15.072 & 15.194 & 15.246\ph \\
      CC4         & 13.920 & 14.007 & 14.052 & 14.076 & 14.196 & 14.308 & 14.359 & 14.388 & 14.304 & 14.331 & 14.413 & 14.439\ph \\
      CCSDTQ      & 13.879 & 13.969 & 14.041 &     \x & 14.182 & 14.200 & 14.310 &     \x & 14.209 & 14.316 & 14.423 &     \x\ph \\
      FCI         & 13.798 & 13.889 & 13.944 & 13.963 & 14.084 & 14.099 & 14.167 & 14.193 & 14.108 & 14.244 & 14.337 & 14.359(1) \\
      \hline
      Mol.        & \mc{8}{c}{Lithium fluoride (\ce{LiF})} & \mc{4}{c}{Beryllium oxide (\ce{BeO})} \\
      State/Conf. & \mc{4}{c}{$1~^{2}\Sigma^-$/$(1\pi)^{-2}(5\sigma)^1$} & \mc{4}{c}{$2~^{2}\Pi$/$(4\sigma)^{-1}(1\pi)^{-1}(5\sigma)^1$} & \mc{4}{c}{$1~^{2}\Sigma^-$/$(1\pi)^{-2}(5\sigma)^1$} \\
      Exp.        & \mc{4}{c}{} & \mc{4}{c}{} & \mc{4}{c}{} \\
      \cline{2-5} \cline{6-9} \cline{10-13}
      CC3         &     \x &     \x &     \x &     \x &     \x &     \x &     \x &     \x\ph &     \x &     \x &     \x &     \x \\
      CCSDT       & 26.917 & 27.177 & 27.738 & 27.945 & 27.545 & 27.810 & 28.345 & 28.559\ph & 15.515 & 15.699 & 16.062 & 16.206 \\
      CC4         & 24.868 & 25.062 & 25.341 &     \x & 25.125 & 25.295 & 25.565 &     \x\ph & 13.215 & 13.376 &     \x &     \x \\
      CCSDTQ      & 25.937 & 26.105 & 26.401 &     \x & 26.464 & 26.632 & 26.900 &     \x\ph & 14.198 & 14.349 & 14.517 &     \x \\
      FCI         & 25.958 & 26.118 & 26.381 &     \x & 26.471 & 26.627 & 26.856 & 27.016(1) & 14.095 & 14.244 & 14.380 &     \x \\
      \hline
      Mol.        & \mc{4}{c}{Beryllium oxide (\ce{BeO})} & \mc{8}{c}{Boron nitride (\ce{BN})} \\
      State/Conf. & \mc{4}{c}{$2~^{2}\Pi$/$(4\sigma)^{-1}(1\pi)^{-1}(5\sigma)^1$} & \mc{4}{c}{$1~^{2}\Sigma^-$/$(1\pi)^{-2}(5\sigma)^1$} & \mc{4}{c}{$1~^{2}\Delta$/$(1\pi)^{-2}(5\sigma)^1$} \\
      Exp.        &  &  &  &  &  &  &  &  &  &  &  &  \\
      \cline{2-5} \cline{6-9} \cline{10-13}
      CC3         &     \x &     \x &     \x &     \x & 13.120 & 13.132 & 13.270 & 13.289 & 13.299 & 13.331 & 13.489 & 13.517\phh \\
      CCSDT       & 17.306 & 17.501 & 17.826 & 17.984 & 13.432 & 13.515 & 13.795 & 13.870 & 13.942 & 14.024 & 14.252 & 14.315\phh \\
      CC4         & 13.361 & 13.558 &     \x &     \x & 12.569 & 12.627 & 12.758 & 12.790 & 13.221 & 13.289 & 13.383 & 13.402\phh \\
      CCSDTQ      & 15.677 & 15.840 & 15.954 &     \x & 12.510 & 12.582 & 12.739 &     \x & 13.244 & 13.324 & 13.431 &     \x\phh \\
      FCI         & 15.455 & 15.616 & 15.683 & 15.805 & 12.393 & 12.463 & 12.600 &     \x & 13.185 & 13.263 & 13.351 & 13.357(21) \\
    \end{tabular}
  \end{ruledtabular}
\end{table*}
\end{squeezetable}

\subsection{12-electron molecules: \ce{LiF}, \ce{BeO}, \ce{BN}, and \ce{C2}}

We now direct our attention toward the 12-electron isoelectronic molecules: \ce{LiF}, \ce{BeO}, \ce{BN}, and \ce{C2}.
These four molecules are quite challenging for theoretical methods as, except for \ce{LiF}, their ground states have a strong multi-reference character. \cite{Bauschlicher_1987,Abrams_2004,Sherrill_2005,Li_2006,Booth_2011,Evangelista_2011b,Gulania_2019,Ammar_2024}
For example, \ce{BN} and \ce{BeO} are among the eight molecules of the $GW$100 set having multiple solution issues at the $GW$ level. \cite{vanSetten_2015,Caruso_2016} (\ce{C2} is not considered in the $GW$100 set but would certainly fall in the same category.)
Another noteworthy observation about these molecules is that their lowest unoccupied molecular orbital has a negative energy, which means that their respective anions are stable.

\ce{LiF} is a relatively non-volatile molecule, and as a result, experimental data became accessible during the second phase of the development of ultraviolet photoelectron spectroscopy. \cite{Berkowitz_1979}
In addition, lithium fluoride vapor is not solely composed of monomers but also includes dimers, trimers, or even tetramers, posing challenges for more precise measurements of the Koopmans states of \ce{LiF}.
Berkowitz \etal~measured the first two IPs using \ce{He(I)} photoelectron spectroscopy: 11.50, 11.67, and \SI{11.94}{\eV} for the $1~^{2}\Pi_{3/2,3/2}$, $1~^{2}\Pi_{3/2,1/2}$, and $1~^{2}\Sigma^+$ states, respectively.
In our study, the spin-orbit coupling is not accounted for. Therefore, we report a single value for the $1~^{2}\Pi$ state, while experimentally two distinct ionization energies are measured.
 
To the best of our knowledge, no gas phase experimental values are available for the three remaining species (see Ref.~\onlinecite{Hamrin_1970} for a study in solid phase).
Nonetheless, they are an interesting playground for theoretical methods due to their multi-reference character.
We start by discussing \ce{BeO} as it has the less pronounced multi-reference character out of these three molecules.
Table \ref{tab:tab4} displays the excitation energies of the two lowest Koopmans states and the first two satellites.
These four states have the same dominant configurations and ordering as the ones of lithium fluoride.
However, the satellite states of \ce{BeO} are much lower in energy than those of \ce{LiF}.
The $2~^{2}\Pi$ state of \ce{BeO} is interesting as it exhibits the largest error of this benchmark set at the CCSDTQ level.
At the CC4 level, the $2~^{2}\Pi$ and $1~^{2}\Sigma^-$ states are drastically underestimated.
This is also the case for the two satellite states of \ce{LiF}, these four states having, by far, the largest CC4 errors of this benchmark set.
They are also hugely underestimated at the CC3 level and, as for CC2 and CCSD, we have not reported these energies as they are not meaningful.
Unfortunately, at this stage, we have no clear explanation for the failure of CC3 and CC4 in \ce{LiF} and \ce{BeO}.

\ce{BN} and \ce{C2} have the strongest multi-reference character among these four molecules. \cite{Ammar_2024}
The ordering of their state differs from the one of \ce{LiF} and \ce{BeO} as their lowest satellite states are below their second IP.
Furthermore, the ordering of the satellites is also different than the two previous molecules.
The first satellite of boron nitride has the same dominant configuration as in the latter two molecules but the second satellite is of $1~^{2}\Delta$ symmetry with a $(1\pi)^{-2}(5\sigma)^1$ dominant process.
This satellite is also found in \ce{C2} but even lower in the energy spectrum as the $1~^{2}\Delta_g$ state is the lowest-energy satellite of the carbon dimer.
Table \ref{tab:tab4} reports two additional FCI satellite transition energies of \ce{C2}.
Note that the satellite transition energies of \ce{BeO}, \ce{BN}, and \ce{C2} are the lowest of the present set.
The three satellite states of the carbon dimer have already been studied by Chatterjee and Sokolov. \cite{Chatterjee_2019,Chatterjee_2020}
In particular, they have shown that ADC(3) performs poorly and does not even predict enough satellite states.
On the other hand, their extension of ADC(2) using a multi-determinantal reference \cite{Sokolov_2018} can predict each state and be in quantitative agreement with FCI. \cite{Chatterjee_2019,Chatterjee_2020}

\begin{squeezetable}
\begin{table*}
  \caption{Valence ionizations and satellite transition energies (in \si{\eV}) of the third-row molecules for various methods and basis sets. 
  AVXZ stands for aug-cc-pVXZ (where X = D, T, and Q). Selected experimental values are also reported.}
  \label{tab:tab5}
  \begin{ruledtabular}
    \begin{tabular}{rccccccccccccc}
      & \mc{4}{c}{Basis} & \mc{4}{c}{Basis} & \mc{4}{c}{Basis} \\
      \cline{2-5} \cline{6-9} \cline{10-13}
      Methods & \mcc{6-31$+$G$^{*}$} & \mcc{AVDZ} & \mcc{AVTZ} & \mcc{AVQZ} & \mcc{6-31$+$G$^{*}$} & \mcc{AVDZ} & \mcc{AVTZ} & \mcc{AVQZ} & \mcc{6-31$+$G$^{*}$} & \mcc{AVDZ} & \mcc{AVTZ} & \mcc{AVQZ}  \\
      \hline
      Mol.        & \mc{12}{c}{Carbon sulfide (\ce{CS})} \\
      State/Conf. & \mc{4}{c}{$1~^{2}\Sigma^+ $/$(7\sigma)^{-1}$} & \mc{4}{c}{$1~^{2}\Pi $/$(2\pi)^{-1}$} & \mc{4}{c}{$~^{2}\Sigma^+ $/$(6\sigma)^{-1}$} \\
      Exp.        & \mc{4}{c}{} & \mc{4}{c}{} & \mc{4}{c}{} \\
      \cline{2-5} \cline{6-9} \cline{10-13}
      CC2         & 10.627 & 10.745 & 10.847 & 10.900 & 12.791 & 12.897 & 13.014\ph & 13.083 & 16.698 & 16.817 & 16.945 & 17.005\ph \\
      CCSD        & 11.245 & 11.402 & 11.553 & 11.609 & 12.726 & 12.883 & 13.000\ph & 13.059 & 16.854 & 16.997 & 17.220 & 17.288\ph \\
      CC3         & 10.949 & 11.186 & 11.325 & 11.377 & 12.553 & 12.766 & 12.880\ph & 12.937 & 18.201 & 18.290 & 18.389 & 18.422\ph \\
      CCSDT       & 10.966 & 11.190 & 11.346 & 11.404 & 12.596 & 12.799 & 12.918\ph & 12.974 & 17.915 & 18.023 & 18.134 & 18.179\ph \\
      CC4         & 10.914 & 11.161 & 11.310 & 11.368 & 12.542 & 12.764 & 12.878\ph & 12.934 & 17.764 & 17.881 & 17.959 & 17.994\ph \\
      CCSDTQ      & 10.920 & 11.166 & 11.316 &     \x & 12.548 & 12.768 & 12.885\ph &     \x & 17.749 & 17.865 & 17.947 &     \x\ph \\
      FCI         & 10.899 & 11.151 & 11.300 & 11.355 & 12.545 & 12.768 & 12.882(1) & 12.936 & 17.723 & 17.844 & 17.920 & 17.958(2) \\
      $G_0W_0$    & 12.092 & 12.119 & 12.378 & 12.523 & 12.602 & 12.679 & 12.907\ph & 13.063 & 17.666 & 17.713 & 17.976 & 18.119\ph \\
      qs$GW$      & 11.369 & 11.589 & 11.775 & 11.880 & 12.498 & 12.705 & 12.852\ph & 12.958 & 17.323 & 17.505 & 17.688 & 17.788\ph \\
      G$_0$F(2)   & 11.109 & 11.152 & 11.292 & 11.371 & 12.696 & 12.774 & 12.923\ph & 13.018 & 16.704 & 16.779 & 16.942 & 17.026\ph \\
      $G_0T_0$    & 11.595 & 11.594 & 11.713 &     \x & 12.479 & 12.505 & 12.609\ph &     \x & 17.422 & 17.455 & 17.591 &     \x\ph \\
      \hline
      Mol.        & \mc{12}{c}{Carbon sulfide (\ce{CS})} \\
      State/Conf. & \mc{4}{c}{$2~^{2}\Sigma^+ $/$(2\pi)^{-1}(7\sigma)^{-1}(3\pi)^1$} & \mc{4}{c}{$3~^{2}\Sigma^+$/$(2\pi)^{-1}(7\sigma)^{-1}(3\pi)^1$} & \mc{4}{c}{$1~^{2}\Delta$/$(2\pi)^{-1}(7\sigma)^{-1}(3\pi)^1$} \\
      Exp.        & \mc{4}{c}{} & \mc{4}{c}{} & \mc{4}{c}{} \\
      \cline{2-5} \cline{6-9} \cline{10-13}
      CC3         & 16.183 & 16.329 & 16.500 & 16.560\ph & 17.358 & 17.491 & 17.558 & 17.584\ph & 17.448 & 17.558 & 17.635 & 17.662\ph \\
      CCSDT       & 15.921 & 16.089 & 16.278 & 16.352\ph & 16.986 & 17.145 & 17.208 & 17.264\ph & 17.039 & 17.178 & 17.266 & 17.319\ph \\
      CC4         & 15.677 & 15.863 & 15.997 & 16.059\ph & 16.628 & 16.799 & 16.773 & 16.807\ph & 16.701 & 16.861 & 16.871 & 16.902\ph \\
      CCSDTQ      & 15.646 & 15.838 & 15.982 &     \x\ph & 16.600 & 16.772 & 16.754 &     \x\ph & 16.678 & 16.840 & 16.858 &     \x\ph \\
      FCI         & 15.604 & 15.803 & 15.935 & 15.996(1) & 16.551 & 16.727 & 16.691 & 16.728(2) & 16.632 & 16.800 & 16.802 & 16.837(2) \\
      \hline
      Mol.        & \mc{8}{c}{Lithium chloride (\ce{LiCl})} & \mc{4}{c}{} \\
      State/Conf. & \mc{4}{c}{$1~^{2}\Pi$/$(2\pi)^{-1}$} & \mc{4}{c}{$1~^{2}\Sigma^+$/$(6\sigma)^{-1}$} & \mc{4}{c}{} \\
      Exp.        & \mc{4}{c}{9.98,10.06 \cite{Patanen_2012}} & \mc{4}{c}{10.77 \cite{Patanen_2012}} & \mc{4}{c}{} \\
      \cline{2-5} \cline{6-9} \cline{10-13}
      CC2         & 9.215 & 9.396 & 9.535 & \z9.639 & \z9.902 & 10.109 & 10.189 & 10.299 &        &        &        &        \\
      CCSD        & 9.604 & 9.830 & 9.956 &  10.067 &  10.327 & 10.566 & 10.637 & 10.757 &  &  &  &  \\
      CC3         & 9.529 & 9.788 & 9.880 & \z9.992 &  10.235 & 10.518 & 10.552 & 10.671 &        &        &        &        \\
      CCSDT       & 9.533 & 9.787 & 9.883 & \z9.993 &  10.241 & 10.517 & 10.554 & 10.673 &  &  &  &  \\
      CC4         & 9.556 & 9.811 & 9.898 &      \x &  10.265 & 10.544 & 10.573 &     \x &        &        &        &        \\
      CCSDTQ      & 9.552 & 9.808 & 9.896 &      \x &  10.261 & 10.540 & 10.570 &     \x &  &  &  &  \\
      FCI         & 9.555 & 9.810 & 9.897 &  10.007 &  10.267 & 10.545 & 10.577 & 10.696 &  &  &  &  \\
      $G_0W_0$    & 9.611 & 9.734 & 9.984 &  10.180 &  10.357 & 10.500 & 10.690 & 10.897 &  &  &  &  \\
      qs$GW$      & 9.574 & 9.808 & 9.947 &  10.086 &  10.307 & 10.569 & 10.642 & 10.789 &  &  &  &  \\
      G$_0$F(2)   & 9.222 & 9.365 & 9.551 & \z9.676 & \z9.916 & 10.083 & 10.210 & 10.341 &  &  &  &  \\
      $G_0T_0$    & 9.567 & 9.619 & 9.761 &      \x &  10.274 & 10.360 & 10.448 &     \x &  &  &  &  \\
      \hline
      Mol.        & \mc{8}{c}{Lithium chloride (\ce{LiCl})} & \mc{4}{c}{Fluorine (\ce{F2})} \\
      State/Conf. & \mc{4}{c}{$2~^{2}\Sigma^+$/$(2\pi)^{-2}(7\sigma)^1$} & \mc{4}{c}{$2~^{2}\Pi$/$(6\sigma)^{-1}(2\pi)^{-1}(7\sigma)^1$} & \mc{4}{c}{$2~^{2}\mathrm{\Sigma}_g^+$/$(1\pi_u)^{-1}(1\pi_g)^{-1}(3\sigma_u)^{1}$} \\
      Exp.        & \mc{4}{c}{} & \mc{4}{c}{} & \mc{4}{c}{} \\
      \cline{2-5} \cline{6-9} \cline{10-13}
      CC3         & 18.447 & 18.788 & 19.156 & 19.328 & 19.055 & 19.402 & 19.717 & 19.897 & 22.465\ph & 22.584\ph & 22.866\ph & 22.934\ph \\
      CCSDT       & 19.582 & 20.043 & 20.508 & 20.729 & 20.304 & 20.745 & 21.168 & 21.397 & 22.293\ph & 22.387\ph & 22.663\ph & 22.758\ph \\
      CC4         & 18.837 & 19.326 & 19.639 &     \x & 19.477 & 19.959 & 20.224 &     \x & 22.050\ph & 22.064\ph & 22.177\ph & 22.234\ph \\
      CCSDTQ      & 18.963 & 19.468 & 19.788 &     \x & 19.645 & 20.141 & 20.413 &     \x & 22.025\ph & 22.038\ph & 22.174\ph &     \x\ph \\
      FCI         & 18.942 & 19.446 & 19.741 & 19.955 & 19.617 & 20.115 & 20.357 & 20.577 & 22.024(2) & 22.039(1) & 22.165(1) & 22.224(4) \\
      \hline
      Mol.        & \mc{12}{c}{Fluorine (\ce{F2})} \\
      State/Conf. & \mc{4}{c}{$1~^{2}\mathrm{\Pi}_g$/$(1\pi_g)^{-1}$} & \mc{4}{c}{$1~^{2}\mathrm{\Pi}_u$/$(1\pi_u)^{-1}$} & \mc{4}{c}{$1~^{2}\mathrm{\Sigma}_g^+$/$(3\sigma_g)^{-1}$} \\
      Exp.        & \mc{4}{c}{15.8 \cite{Zheng_1996}} & \mc{4}{c}{18.9 \cite{Zheng_1996}} & \mc{4}{c}{20.9 \cite{Zheng_1996}}  \\
      \cline{2-5} \cline{6-9} \cline{10-13}
      CC2         & 13.903 & 14.001 & 14.145 & 14.233 & 17.050 & 17.122 & 17.224 & 17.297\ph & 20.325 & 20.458 & 20.522\ph & 20.604 \\
      CCSD        & 15.279 & 15.405 & 15.616 & 15.722 & 18.633 & 18.753 & 18.946 & 19.047\ph & 21.068 & 21.155 & 21.174\ph & 21.241 \\
      CC3         & 15.574 & 15.646 & 15.746 & 15.825 & 18.786 & 18.847 & 18.924 & 18.994\ph & 21.091 & 21.155 & 21.130\ph & 21.182 \\
      CCSDT       & 15.529 & 15.594 & 15.688 & 15.767 & 18.745 & 18.797 & 18.865 & 18.936\ph & 21.048 & 21.109 & 21.094\ph & 21.148 \\
      CC4         & 15.555 & 15.621 & 15.701 & 15.804 & 18.746 & 18.797 & 18.864 & 18.941\ph & 21.086 & 21.142 & 21.146\ph & 21.171 \\
      CCSDTQ      & 15.559 & 15.623 & 15.725 &     \x & 18.754 & 18.803 & 18.874 &     \x\ph & 21.077 & 21.132 & 21.106\ph &     \x \\
      FCI         & 15.564 & 15.628 & 15.729 & 15.808 & 18.758 & 18.807 & 18.874 & 18.943(1) & 21.077 & 21.132 & 21.100(2) & 21.149 \\
      $G_0W_0$    & 15.763 & 15.964 & 16.334 & 16.559 & 19.423 & 19.589 & 19.902 & 20.104\ph & 20.434 & 20.625 & 20.836\ph & 21.029 \\
      qs$GW$      & 15.927 & 16.016 & 16.229 & 16.368 & 19.524 & 19.585 & 19.752 & 19.875\ph & 20.855 & 20.934 & 21.000\ph & 21.114 \\
      G$_0$F(2)   & 13.809 & 13.960 & 14.194 & 14.328 & 16.965 & 17.087 & 17.275 & 17.389\ph & 20.137 & 20.308 & 20.420\ph & 20.534 \\
      $G_0T_0$    & 15.026 & 15.157 & 15.401 &     \x & 18.615 & 18.724 & 18.928 &     \x\ph & 19.934 & 20.083 & 20.190\ph &     \x \\
      \end{tabular}
  \end{ruledtabular}
\end{table*}
\end{squeezetable}

\subsection{Third-row molecules: \ce{CS}, \ce{Ar}, \ce{HCl}, \ce{H2S}, \ce{PH3}, \ce{SiH4}, and \ce{LiCl}}

The molecules examined in this subsection have been obtained by substituting a second-row atom with its third-row analog in some of the molecules discussed above.
These molecules with more diffuse density have their ionization shifted towards zero with respect to their second-row counterparts (see Tables \ref{tab:tab5}, \ref{tab:tab6}, and \ref{tab:tab7}).
Consequently, the breakdown of the orbital picture occurs at lower energy, \cite{Schirmer_1978a} which has been of interest historically as it allowed measuring spectra featuring such intricate structures more easily.

The first molecule considered in this subsection is carbon sulfide.
In 1972, two independent studies measured its photoelectron spectrum up to \SI{20}{\eV}. \cite{Frost_1972,Jonathan_1972}
One can clearly distinguish four well-defined peaks in this energy range.
While the assignment of the two lowest peaks is straightforward, \ie~IPs associated with the two outermost orbitals, the interpretation of the other two remained elusive for several years.
Thanks to theoretical studies performed several years later, it became clear that the third peak is due to a multi-particle process while the fourth one is associated with an electron detachment from the orbital $6\sigma$. \cite{Schirmer_1978a,Chatterjee_2019,Chatterjee_2020,Moitra_2022}
Note that, as explained by Schirmer \etal, \cite{Schirmer_1978a} one has to be particularly careful when labeling the third peak as a satellite because its FCI vector has a coefficient of 0.49 on the 1h configuration $(6\sigma)^{-1}$ and of 0.40 on the 2h1p configuration $(2\pi)^{-1}(7\sigma)^{-1}(3\pi)^1$ (in 6-31+G* basis set).
This is yet another example of a strong configuration mixing.

Despite this, the $2~^{2}\Sigma^+$ state is classified as a satellite in Table \ref{tab:tab5}.
Indeed, higher in energy there is another FCI solution of $~^{2}\Sigma^+$ symmetry with an even larger weight, 0.63, on the 1h determinant $(6\sigma)^{-1}$ and coefficients smaller than 0.31 on the 2h1p determinants.
In addition, the third peak has a pronounced vibrational structure while the fourth peak is sharp like the $(7\sigma)^{-1}$ one. \cite{Jonathan_1972}
This is why the higher $~^{2}\Sigma^+$ state is classified as the third single ionization in Table \ref{tab:tab5}. 
Two other satellite states, $3~^{2}\Sigma^+$ and $1~^{2}\Delta$, that are not visible on the experimental spectrum, are reported in Table \ref{tab:tab5}.

Next, we consider lithium chloride and compare it with its second-row analog, lithium fluoride.
The experimental challenges outlined earlier for \ce{LiF} are similar for \ce{LiCl}.
Experimental values have first been reported independently by two groups in 1979, \cite{Potts_1979,Berkowitz_1979} and revised values, measured by \ce{He(I)} spectroscopy, have been published recently, \cite{Patanen_2012} and are reported in Table \ref{tab:tab5}.
The FCI results predict two close-lying satellites around \SI{20}{\eV}.
Unfortunately, the experimental studies mentioned above do not probe this energy range.
The ADC(3) calculations of Tomasello \etal~also predict two satellite lines around \SI{21.5}{\eV}. \cite{Tomasello_1990}

Finally, we examine the 18-electron isoelectronic hydrides as analogs to the 10-electron series discussed in Subsec.~\ref{sec:10_elec}.
Historically, it was quickly realized that the satellite structure of \ce{H2S} is significantly more complex than the one of \ce{H2O}. \cite{Domcke_1978}
This intricate structure can be observed in the electron momentum spectrum of French \etal \cite{French_1988}
They recorded a first very weak satellite at \SI{19.63}{\eV} which is in agreement with earlier measurements \cite{Cook_1980} as well as photoelectron spectrum measured using synchrotron radiation. \cite{Adam_1987}
Several years later, extensive SAC-CI results have been reported and show qualitative agreement with experiments. \cite{Ehara_2001} (See also earlier calculations from Refs.~\onlinecite{Wasada_1989,Lisini_1991}.)
However, the FCI results (see Table~\ref{tab:tab7}) exhibit some significant difference with the SAC-CI results of Ehara \etal~because the $2~^{2}\mathrm{A}_1$ satellite has a lower energy than the $2~^{2}\mathrm{B}_2$ state.
In addition, the FCI transition energy associated with the $2~^{2}\mathrm{B}_1$ state is \SI{2}{\eV} lower than the one computed in Ref.~\onlinecite{Ehara_2001}.
The $2~^{2}\mathrm{A}_1$ state is known to be the one observed at \SI{19.63}{\eV} in Ref.~\onlinecite{French_1988} and is sometimes referred to as a shake-down state as it  ``borrows'' intensity from the higher-lying $(4a_1)^{-1}$ ionization. \cite{Chipman_1978,Adam_1987,Ehara_2001}
Our FCI estimate for this state is \SI{18.745}{\eV} while the SAC-CI energy of Ehara and collaborators is \SI{20.00}{\eV}. \cite{Ehara_2001}
In this specific scenario, calculating the adiabatic transition energy related to this state would undoubtedly provide a more faithful comparison with the experimental result.

For \ce{PH3}, there is one satellite of symmetry E that is analog to the two $2~^{2}\mathrm{E}$ state of \ce{NH3}.
However, in the case of phosphine, there is no analog for the $2~^{2}\mathrm{A}_1$ satellite of ammonia.
This is in agreement with the SAC-CI results of Ishida \etal~who also found a single satellite below the $(4a_1)^{-1}$ ionization threshold. \cite{Ishida_2002}

Two FCI states, with symmetry $^{2}\Sigma^+$ and $^{2}\Delta$, are reported for \ce{HCl}.
These states are analog to the \ce{HF} satellites reported in Table \ref{tab:tab1} although they have significantly lower energies in \ce{HCl}.
The satellite structure between \SI{20}{\eV} and the $(4\sigma)^{-1}$ ionization around \SI{26}{\eV} is notably intricate. \cite{Yencha_1998}
Additionally, this structure is characterized by weak signals and some of its features were even not observed in previous studies performed at a lower level of theory. \cite{Adam_1986,Svensson_1988a,Edvardsson_1995}
The assignment of the various peaks in this energy range is beyond the scope of this work.
Yet, one can mention that the first FCI satellite is in qualitative agreement with the first satellite peak measured by synchrotron radiation spectroscopy at \SI{21.57}{\eV}. \cite{Yencha_1998}

The lowest-energy satellite of argon, which is the analog of the $2~^{2}P$ satellite state of neon, is also reported.
This satellite has been observed experimentally by Kikas \etal and is also reported in Table~\ref{tab:tab7}. \cite{Kikas_1996}
The agreement of the FCI value with the experimental one is definitely not as good as for neon.
On the other hand, CC3 provides an excellent approximation of the $2~^{2}P$ satellite state of argon which is not the case for neon (see Table~\ref{tab:tab1}).


\subsection{Miscellaneous molecules: \ce{F2}, \ce{CO2}, \ce{CH2O}, and \ce{BH3}}

In this last subset, a few miscellaneous molecules are considered.
First, the \ce{F2} molecule is of interest as it is isoelectronic to \ce{SiH4}, \ce{PH3}, \ce{H2S}, \ce{HCl} and \ce{Ar}.
Due to the absence of third-row atoms in \ce{F2}, Cederbaum \etal~observed that there is no breakdown of the orbital picture, as observed in the 18-electron hydride series. \cite{Cederbaum_1977}
Three IPs and one satellite transition energy of fluorine are reported in Table \ref{tab:tab5}.
As documented in Ref.~\onlinecite{Cederbaum_1977}, these four states have a clear dominant configuration in their corresponding FCI vectors.
The $2~^{2}\mathrm{\Sigma}_g^+$ satellite is not observed in photoionization \cite{Guyon_1976,Bieri_1980} or in electron impact spectra. \cite{Zheng_1996}
However, a satellite state with the same symmetry and similar energy has been computed using ADC(4) and multi-reference CI. \cite{Zheng_1996}

Carbon dioxide and formaldehyde are two small organic molecules that have been widely studied both experimentally and theoretically.
Experimental studies have shown that \ce{CO2} first measurable satellite peak is around \SI{22}{\eV}, \cite{Potts_1974,Domcke_1979,Roy_1986,Tian_2012} while four ionization peaks are observed below \SI{20}{\eV}.
Tian's experimental values for these IPs, measured by electron momentum spectroscopy, are reported alongside our FCI estimates in Table \ref{tab:tab8}. \cite{Tian_2012}
These IPs have already been computed at various levels of theory such as CI, \cite{Roy_1986} ADC, \cite{Domcke_1979,Chatterjee_2019,Chatterjee_2020} SAC-CI, \cite{Nakatsuji_1983,Tian_2012} CC, \cite{Ranasinghe_2019} and even FCI. \cite{Chatterjee_2019,Chatterjee_2020}

The spectrum of formaldehyde is slightly harder to interpret.
The electron momentum spectrum displays a shark peak at \SI{10.9}{\eV} as well as a broad band between 12 and \SI{18}{\eV}. \cite{Bawagan_1988a}
On the other hand, one can observe four different peaks below \SI{17}{\eV} in the corresponding photoionization spectra. \cite{vonNiessen_1980,Hochlaf_2005}
These peaks clearly correspond to ionizations from the four outermost orbitals (see Table \ref{tab:tab8}) and have already been computed using both wave function and Green's function methods. \cite{vonNiessen_1980,Morrison_1992,Musial_2003a,Ranasinghe_2019,Chatterjee_2019,Chatterjee_2020}
Hochlaf \etal~also mention a very weak band around \SI{18}{\eV} assigned as a satellite of $^{2}\mathrm{B}_2$ symmetry, \cite{Hochlaf_2005} which can also be observed in Ref.~\onlinecite{vonNiessen_1980}. 
This is in nice agreement with the $3~^{2}\mathrm{B}_2$ FCI satellite state.
There is an additional $2~^{2}\mathrm{B}_1$ satellite state with slightly lower energy than the previous one.

Finally, a small boron hydride is considered as we have seen above that boron-containing molecules such as \ce{BN} are quite challenging.
To the best of our knowledge, experimental results on \ce{BH3} are quite scarce but the principal IP has been measured using mass spectroscopy in the 60's. \cite{Fehlner_1964,Wilson_1967}
These two research groups reported quite different values of \SI{12.32}{\eV} \cite{Fehlner_1964} and \SI{11.4}{\eV}. \cite{Wilson_1967}
The FCI values, presented in Table \ref{tab:tab8}, is closer to the first one, corroborating the findings of Tian \etal, who computed similar values using propagator-based methods as well as CCSD(T). \cite{Tian_2005}
Finally, two FCI satellite transition energies of \ce{BH3} are reported in Table \ref{tab:tab8}.

\subsection{Global statistics}
\label{sec:stat}

\begin{figure}
  \centering
  \includegraphics[width=\linewidth]{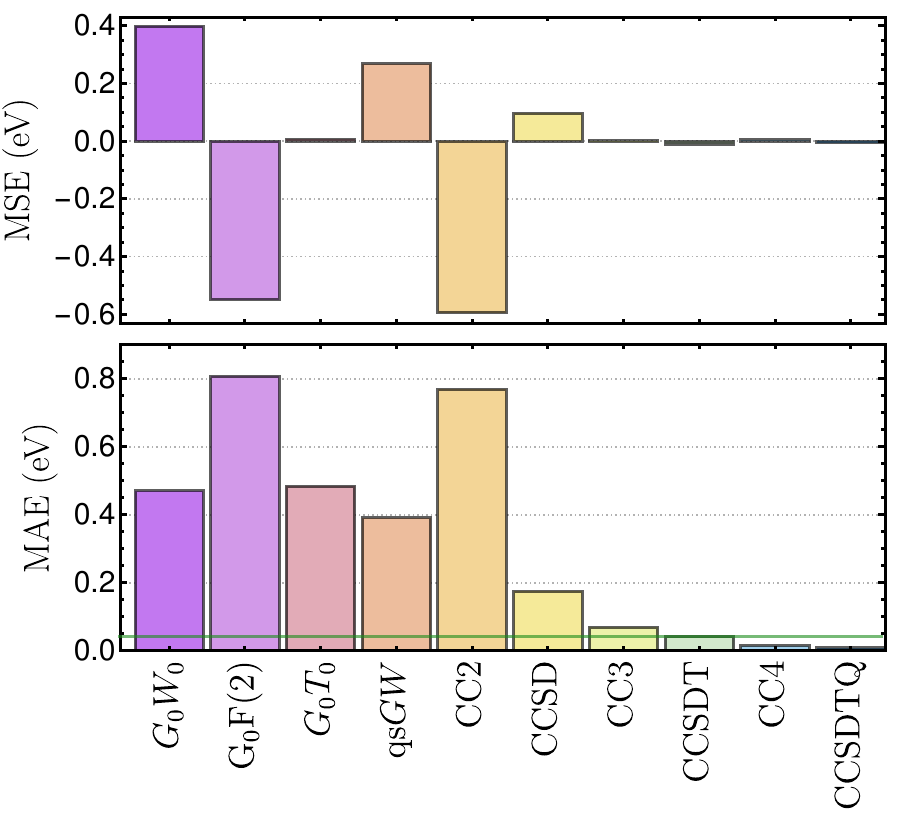}
  \caption{Mean signed error (MSE) [upper panel] and mean absolute error (MSE) [lower panel] with respect to FCI of the various methods considered in this work. These errors are computed for the 58 IPs of this set in the aug-cc-pVTZ basis set.}
    \label{fig:StatIP}
\end{figure}

\begin{table}
  \caption{Mean absolute error (MAE), mean signed error (MSE), root mean square error (RMSE), standard deviation error (SDE), minimum and maximum errors (in \si{\eV}) with respect to FCI of the various methods considered in this work. These descriptors are computed for the 58 IPs of this set in the aug-cc-pVTZ basis set. The $\Delta$CCSD(T) statistical descriptors correspond only to the 23 principal IPs.}
   \label{tab:tab6}
  \begin{ruledtabular}
    \begin{tabular}{lcccccc}
      Methods         &  MAE  &  MSE    &  RMSE &  SDE  &  Min   &  Max  \\
      \hline              
      CC2             & 0.769 &  -0.594 & 0.940 & 0.735 & -2.207 & 1.565 \\
      CCSD            & 0.175 & \z0.097 & 0.280 & 0.265 & -0.700 & 1.075 \\
      CC3             & 0.070 & \z0.001 & 0.125 & 0.126 & -0.395 & 0.469 \\
      CCSDT           & 0.041 &  -0.010 & 0.057 & 0.057 & -0.140 & 0.214 \\
      CC4             & 0.015 & \z0.005 & 0.027 & 0.027 & -0.078 & 0.118 \\
      CCSDTQ          & 0.010 &  -0.005 & 0.013 & 0.012 & -0.049 & 0.027 \\
      \hline                                                        
      $G_0W_0$        & 0.470 & \z0.399 & 0.664 & 0.535 & -0.504 & 2.053 \\
      qs$GW$          & 0.391 & \z0.268 & 0.559 & 0.494 & -1.348 & 1.747 \\
      G$_0$F(2)       & 0.807 &  -0.550 & 0.987 & 0.827 & -2.336 & 1.623 \\
      $G_0T_0$        & 0.485 & \z0.007 & 0.752 & 0.758 & -1.169 & 2.959 \\
      \hline
      $\Delta$CCSD(T) & 0.021 & \z0.016 & 0.037 & 0.035 & -0.020 & 0.120 \\
    \end{tabular}
  \end{ruledtabular}
\end{table}

Finally, after discussing each molecule individually, the statistics over the whole set are reported and discussed in this subsection.
Figure \ref{fig:StatIP} displays the mean sign error (MSE) and MAE of the various methods considered in this study with respect to the new FCI references.
These statistical errors have been computed for the 58 IPs in the aug-cc-pVTZ basis set.
Several other statistical descriptors are also reported in Table \ref{tab:tab6}.

The CC hierarchy (CCSD, CCSDT, and CCSDTQ) behaves as expected, \ie~being more and more accurate as the rank of the excitation is increased.
Chemical accuracy (\ie, error below \SI{0.043}{\eV} as represented by the horizontal green line in the lower panel Fig.~\ref{fig:StatIP}) is reached at the CCSDT level.
The least expensive CC2 method does not perform well for IPs as already observed previously. \cite{Walz_2016,Dutta_2018,PauleyParan_2024}
This has been attributed to the same underlying issue observed in CC2 for Rydberg \cite{Kannar_2017} and charge-transfer excited states.\cite{Kozma_2020}

Figure \ref{fig:StatIP} also shows that CC3 and CC4 are good approximations, for IPs, of their respective parents, CCSDT and CCSDTQ.
This could have been expected for CC3 as it is known to be a good approximation of CCSDT for Rydberg excited states. \cite{Kannar_2017}
In addition, these four methods have very small MSEs and do not, on average, underestimate (as CC2) or overestimate (as CCSD) the IPs.
Therefore, implementations of IP-EOM-CC3 and IP-EOM-CC4 would be valuable to lower the cost of the present implementation based on EE-EOM.
CC3 and CC4 could be certainly employed as reference methods for larger molecular systems. \cite{Patanen_2021,Loos_2021a,Loos_2022b}

For the sake of completeness, we also report in Table \ref{tab:tab6} the statistical descriptors for the propagator methods. However, their trends are now well-known. \cite{Zhang_2017,Bruneval_2021,Monino_2023,Marie_2023}
The $G_0T_0$ MAE is very close to the $G_0W_0$ one whereas the second-Born approximation exhibits significantly poorer performance. \cite{Zhang_2017,Bruneval_2021,Monino_2023}
The self-consistent qs$GW$ slightly mitigates the error compared to the one-shot $GW$ version.
It is also interesting to note that GF2 results are very close to those of CC2.
This could have been expected as GF2 is equivalent to ADC(2) \cite{Schirmer_2018,Backhouse_2021,Monino_2023} and the latter is closely related to the CC2 approximation. \cite{Hattig_2005c}

Finally, the principal IP of the 23 molecules considered so far have been computed at the $\Delta$CCSD(T) in the aug-cc-pVTZ basis set (see \SupInf).
This method has been used as the reference for the $GW$100 dataset and it is interesting to benchmark it now that we have access to FCI references. \cite{Krause_2015,Caruso_2016,Govoni_2018,Bruneval_2021}
The last line of Table \ref{tab:tab6} reports the corresponding statistical descriptor.
In particular, its MAE and MSE of \SI{0.021}{\eV} and \SI{0.016}{\eV}, respectively, show that the state-specific $\Delta$CCSD(T) method can indeed be employed as a reference.

\begin{figure*}
  \centering
  \includegraphics[width=\linewidth]{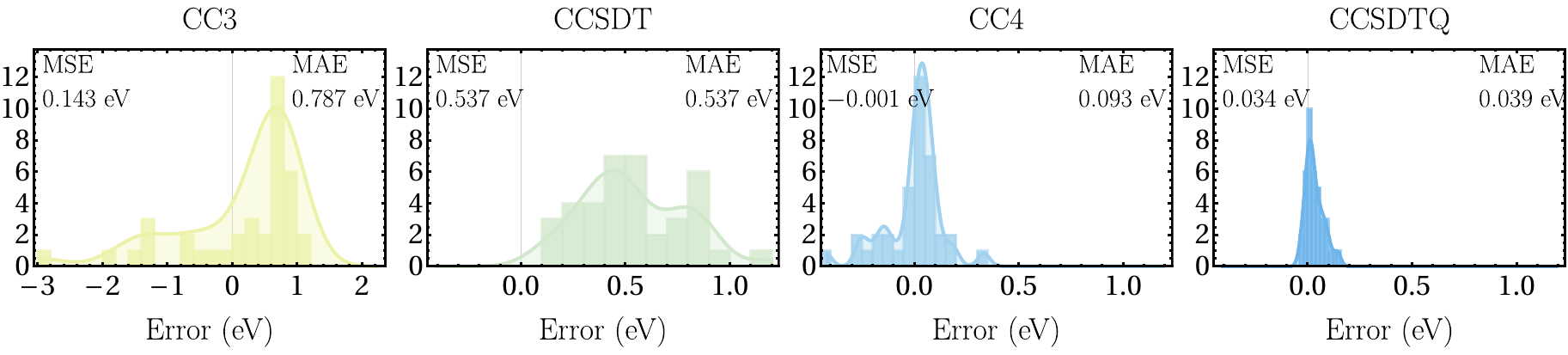}
  \caption{Distribution of the errors with respect to FCI of the various methods considered in this work. These errors are computed for the 36 satellites of this set in the aug-cc-pVTZ basis set. The satellites of \ce{LiF} and \ce{BeO} have been excluded (see main text). Note the different scale of the horizontal axis in the leftmost plot.}
    \label{fig:StatSat}
\end{figure*}

Figure \ref{fig:StatSat} shows the distribution of the errors associated with the satellite transitions computed with CC methods including at least triple excitations, namely, CC3, CCSDT, CC4, and CCSDTQ. 
The corresponding statistical descriptors are reported in Table \ref{tab:tab7}.
The MAE of CCSDTQ is \SI{0.039}{\eV}, \ie~just below chemical accuracy, while CC4 and its approximate treatment of quadruples achieve a \SI{0.093}{\eV} MAE.
Interestingly, while CC4 absolute errors are, on average, larger than CCSDTQ, its MSE is closer to zero.
CC3 and CCSDT have MAEs of \SI{0.787}{\eV} and \SI{0.537}{\eV}, respectively, and once again CC3 has a lower MSE than its parent method (\SI{0.143}{\eV} and \SI{0.537}{\eV}).
Hence, methods accounting for triple excitations, even fully, should be used with care for satellites.
It is also interesting to note that these MAEs align very well with those computed for double excitations, as reported in Ref.~\onlinecite{Loos_2019}.

\begin{table}
  \caption{Mean absolute error (MAE), mean signed error (MSE), root mean square error (RMSE), standard deviation error (SDE), minimum and maximum errors (in \si{\eV}) with respect to FCI of the various methods considered in this work. These descriptors are computed for the 36 satellites of this set in the aug-cc-pVTZ basis set. The satellites of \ce{LiF} and \ce{BeO} have been excluded (see main text).}
  \label{tab:tab7}
  \begin{ruledtabular}
    \begin{tabular}{lcccccc}
      Methods     &  MAE  &  MSE  & RMSE  &  SDE  &   Min   &  Max  \\
      \hline
      CC3         & 0.787 &  0.143 & 0.936 & 0.937 &  -2.907 & 1.098 \\
      CCSDT       & 0.537 &  0.537 & 0.590 & 0.248 & \z0.104 & 1.195 \\
      CC4         & 0.093 & -0.001 & 0.134 & 0.136 &  -0.423 & 0.333 \\
      CCSDTQ      & 0.039 &  0.034 & 0.054 & 0.043 &  -0.030 & 0.143 \\
    \end{tabular}
  \end{ruledtabular}
\end{table}

\squeezetable
\begingroup
\begin{table*}
  \caption{Valence ionizations and satellite transition energies (in \si{\eV}) of the 18-electron series for various methods and basis sets. 
  AVXZ stands for aug-cc-pVXZ (where X = D, T, and Q). Selected experimental values are also reported.}
  \label{tab:tab8}
  \begin{ruledtabular}
    \begin{tabular}{rccccccccccccc}
      & \mc{4}{c}{Basis} & \mc{4}{c}{Basis} & \mc{4}{c}{Basis} \\
      \cline{2-5} \cline{6-9} \cline{10-13}
      Methods & \mcc{6-31$+$G$^{*}$} & \mcc{AVDZ} & \mcc{AVTZ} & \mcc{AVQZ} & \mcc{6-31$+$G$^{*}$} & \mcc{AVDZ} & \mcc{AVTZ} & \mcc{AVQZ} & \mcc{6-31$+$G$^{*}$} & \mcc{AVDZ} & \mcc{AVTZ} & \mcc{AVQZ}  \\
      \hline
      Mol.        & \mc{12}{c}{ (\ce{H2S})} \\
      State/Conf. & \mc{4}{c}{$1~^{2}\mathrm{B}_1$/$(2b_1)^{-1}$} & \mc{4}{c}{$1~^{2}\mathrm{A}_1$/$(5a_1)^{-1}$} & \mc{4}{c}{$1~^{2}\mathrm{B}_2$/$(2b_2)^{-1}$} \\
      Exp.        &  \mc{4}{c}{10.5 \cite{French_1988}} & \mc{4}{c}{13.1 \cite{French_1988}} & \mc{4}{c}{15.6 \cite{French_1988}} \\
      \cline{2-5} \cline{6-9} \cline{10-13}
      CC2         & \z9.699 & \z9.953 & 10.156 & 10.228 & 12.833 & 13.004 & 13.155 & 13.220 & 15.393 & 15.368 & 15.464 & 15.521 \\
      CCSD        & \z9.908 &  10.208 & 10.421 & 10.490 & 13.057 & 13.280 & 13.440 & 13.497 & 15.540 & 15.591 & 15.684 & 15.734 \\
      CC3         & \z9.846 &  10.189 & 10.388 & 10.454 & 12.985 & 13.259 & 13.407 & 13.459 & 15.475 & 15.567 & 15.635 & 15.680 \\
      CCSDT       & \z9.853 &  10.190 & 10.390 & 10.455 & 12.985 & 13.253 & 13.399 & 13.450 & 15.460 & 15.552 & 15.619 & 15.664 \\
      CC4         & \z9.855 &  10.199 & 10.394 & 10.458 & 12.990 & 13.265 & 13.407 & 13.456 & 15.458 & 15.560 & 15.624 & 15.668 \\
      CCSDTQ      & \z9.855 &  10.199 & 10.393 &     \x & 12.990 & 13.265 & 13.406 &     \x & 15.457 & 15.558 & 15.622 &     \x \\
      FCI         & \z9.855 &  10.199 & 10.393 & 10.456 & 12.992 & 13.268 & 13.411 & 13.460 & 15.459 & 15.562 & 15.627 & 15.672 \\
      $G_0W_0$    &  10.034 &  10.172 & 10.500 & 10.660 & 13.212 & 13.339 & 13.614 & 13.758 & 15.684 & 15.698 & 15.928 & 16.058 \\
      qs$GW$      & \z9.890 &  10.201 & 10.445 & 10.549 & 13.048 & 13.337 & 13.530 & 13.619 & 15.524 & 15.651 & 15.785 & 15.868 \\
      G$_0$F(2)   & \z9.747 & \z9.940 & 10.180 & 10.282 & 12.887 & 13.005 & 13.188 & 13.278 & 15.420 & 15.364 & 15.494 & 15.569 \\
      $G_0T_0$    & \z9.927 &  10.011 & 10.188 &     \x & 13.068 & 13.112 & 13.250 &     \x & 15.588 & 15.477 & 15.582 &     \x \\
      \hline
      Mol.        & \mc{12}{c}{ (\ce{PH3})} \\
      State/Conf. & \mc{4}{c}{$1~^{2}\mathrm{A}_1$/$(5a_1)^{-1}$} & \mc{4}{c}{$1~^{2}\mathrm{E}$/$(2e_g)^{-1}$} &  \mc{4}{c}{$2~^{2}\mathrm{A}_1$/$(4a_1)^{-1}$} \\
      Exp.        & \mc{4}{c}{10.85 \cite{Ishida_2002}} & \mc{4}{c}{16.4 \cite{Ishida_2002}} & \mc{4}{c}{27.6 \cite{Ishida_2002}} \\
      \cline{2-5} \cline{6-9} \cline{10-13}
      CC2         & 10.030 & 10.219 & 10.392 & 10.444 & 13.463 & 13.493 & 13.638 & 13.689 & 21.308 & 21.046 & 21.079 & 21.106 \\
      CCSD        & 10.198 & 10.448 & 10.623 & 10.662 & 13.521 & 13.637 & 13.784 & 13.825 & 20.144 & 20.221 & 20.361 & 20.396 \\
      CC3         & 10.130 & 10.431 & 10.599 & 10.634 & 13.464 & 13.615 & 13.744 & 13.779 & 19.528 & 19.643 & 19.739 & 19.763 \\
      CCSDT       & 10.129 & 10.431 & 10.599 & 10.634 & 13.454 & 13.606 & 13.734 & 13.769 & 19.317 & 19.487 & 19.583 & 19.610 \\
      CC4         & 10.127 & 10.436 & 10.600 &     \x & 13.452 & 13.611 & 13.737 &     \x & 19.263 & 19.446 & 19.520 &     \x \\
      CCSDTQ      & 10.127 & 10.437 & 10.600 &     \x & 13.451 & 13.611 & 13.737 &     \x & 19.259 & 19.442 & 19.511 &     \x \\
      FCI         & 10.126 & 10.436 & 10.596 & 10.628 & 13.456 & 13.615 & 13.745 & 13.784 & 19.261 & 19.445 & 19.514 & 19.537 \\
      $G_0W_0$    & 10.365 & 10.497 & 10.787 & 10.911 & 13.717 & 13.831 & 14.099 & 14.214 & 21.505 & 21.378 & 21.567 & 21.673 \\
      qs$GW$      & 10.190 & 10.506 & 10.725 & 10.802 & 13.545 & 13.761 & 13.954 & 14.027 & 21.128 & 21.155 & 21.261 & 21.322 \\
      G$_0$F(2)   & 10.095 & 10.223 & 10.428 & 10.508 & 13.495 & 13.503 & 13.675 & 13.741 & 21.354 & 21.067 & 21.137 & 21.184 \\
      $G_0T_0$    & 10.186 & 10.236 & 10.387 &     \x & 13.665 & 13.619 & 13.759 &     \x & 22.142 & 21.976 & 22.040 &     \x \\
      \hline
      Mol.        & \mc{8}{c}{(\ce{SiH4})} & \mc{4}{c}{(\ce{HCl})} \\
      State/Conf. & \mc{4}{c}{$1~^{2}\mathrm{T}_2$/$(2t_2)^{-1}$}  & \mc{4}{c}{$1~^{2}\mathrm{A}_1$/$(3a_1)^{-1}$} & \mc{4}{c}{$1~^{2}\Pi$/$(1\pi)^{-1}$} \\
      Exp.        & \mc{4}{c}{12.8 \cite{Clark_1989a}} & \mc{4}{c}{18.2 \cite{Clark_1989a}} & \mc{4}{c}{12.745/12.830 \cite{Yencha_1998}} \\
      \cline{2-5} \cline{6-9} \cline{10-13}
      CC2         & 12.515 & 12.653 & 12.802 & 12.848 & 18.704 & 18.759 & 18.823 & 18.846\ph & 11.990 & 12.230 & 12.401 & 12.491 \\
      CCSD        & 12.477 & 12.701 & 12.844 & 12.877 & 18.266 & 18.379 & 18.483 & 18.505\ph & 12.250 & 12.538 & 12.712 & 12.812 \\
      CC3         & 12.417 & 12.681 & 12.806 & 12.832 & 18.109 & 18.236 & 18.316 & 18.331\ph & 12.186 & 12.529 & 12.672 & 12.771 \\
      CCSDT       & 12.407 & 12.673 & 12.794 & 12.820 & 18.046 & 18.172 & 18.247 & 18.262\ph & 12.190 & 12.524 & 12.667 & 12.764 \\
      CC4         & 12.404 & 12.676 & 12.795 &     \x & 18.033 & 18.171 & 18.241 &     \x\ph & 12.199 & 12.539 & 12.678 & 12.773 \\
      CCSDTQ      & 12.404 & 12.677 & 12.795 &     \x & 18.030 & 18.170 & 18.238 &     \x\ph & 12.199 & 12.538 & 12.676 &     \x \\
      FCI         & 12.403 & 12.676 & 12.793 & 12.818 & 18.031 & 18.173 & 18.240 & 18.258(2) & 12.199 & 12.539 & 12.676 & 12.770 \\
      $G_0W_0$    & 12.716 & 12.960 & 13.214 & 13.312 & 18.701 & 18.824 & 19.007 & 19.097\ph & 12.343 & 12.488 & 12.778 & 12.967 \\
      qs$GW$      & 12.541 & 12.906 & 13.107 & 13.170 & 18.422 & 18.700 & 18.828 & 18.879\ph & 12.224 & 12.539 & 12.723 & 12.855 \\
      G$_0$F(2)   & 12.550 & 12.662 & 12.830 & 12.886 & 18.742 & 18.768 & 18.855 & 18.899\ph & 12.022 & 12.200 & 12.415 & 12.535 \\
      $G_0T_0$    & 12.712 & 12.762 & 12.891 &     \x & 19.119 & 19.156 & 19.204 &     \x\ph & 12.273 & 12.347 & 12.507 &     \x \\
      \hline
      Mol.        & \mc{4}{c}{ (\ce{HCl})} & \mc{8}{c}{Argon (\ce{Ar})} \\
      State/Conf. & \mc{4}{c}{$1~^{2}\Sigma^+$/$(5\sigma)^{-1}$} & \mc{4}{c}{$1~^{2}\mathrm{P}$/$(3p)^{-1}$} & \mc{4}{c}{$1~^{2}\mathrm{S}$/$(3s)^{-1}$} \\
      Exp.        & \mc{4}{c}{16.270 \cite{Yencha_1998}} & \mc{4}{c}{15.8\cite{Svensson_1988}} & \mc{4}{c}{29.24\cite{Svensson_1988}} \\
      \cline{2-5} \cline{6-9} \cline{10-13}
      CC2         & 16.245 & 16.329 & 16.438 & 16.512 & 15.055 & 15.236 & 15.379 & 15.486 & 31.123 & 30.459 & 30.439 & 30.437 \\
      CCSD        & 16.467 & 16.599 & 16.708 & 16.782 & 15.308 & 15.534 & 15.672 & 15.802 & 30.353 & 29.838 & 29.881 & 29.907 \\
      CC3         & 16.395 & 16.577 & 16.658 & 16.730 & 15.230 & 15.525 & 15.614 & 15.745 & 30.255 & 29.437 & 29.238 & 29.227 \\
      CCSDT       & 16.388 & 16.564 & 16.644 & 16.715 & 15.228 & 15.514 & 15.606 & 15.735 & 30.261 & 29.445 & 29.214 & 29.216 \\
      CC4         & 16.394 & 16.577 & 16.653 & 16.723 & 15.238 & 15.531 & 15.616 & 15.743 & 30.263 & 29.442 & 29.186 & 29.179 \\
      CCSDTQ      & 16.393 & 16.576 & 16.651 &     \x & 15.237 & 15.529 & 15.614 & 15.740 & 30.261 & 29.442 & 29.182 & 29.175 \\
      FCI         & 16.396 & 16.579 & 16.657 & 16.728 & 15.237 & 15.529 & 15.613 & 15.739 & 30.265 & 29.449 & 29.188 & 29.182 \\
      $G_0W_0$    & 16.574 & 16.635 & 16.872 & 17.031 & 15.333 & 15.458 & 15.711 & 15.926 & 31.226 & 30.759 & 31.089 & 31.224 \\
      qs$GW$      & 16.417 & 16.619 & 16.752 & 16.861 & 15.232 & 15.507 & 15.633 & 15.794 & 30.858 & 30.576 & 30.681 & 30.751 \\
      G$_0$F(2)   & 16.286 & 16.325 & 16.469 & 16.564 & 15.072 & 15.196 & 15.387 & 15.523 & 31.130 & 30.399 & 30.426 & 30.427 \\
      $G_0T_0$    & 16.461 & 16.436 & 16.548 &     \x & 15.326 & 15.368 & 15.511 &     \x & 32.255 & 32.007 & 32.147 &     \x \\
      \hline
      Mol.        & \mc{12}{c}{ (\ce{H2S})} \\
      State/Conf. & \mc{4}{c}{$2~^{2}\mathrm{A}_1$/$(2b_1)^{-2}(6a_1)^1$} & \mc{4}{c}{$2~^{2}\mathrm{B}_2$/$(2b_1)^{-2}(3b_2)^1$} & \mc{4}{c}{$2~^{2}\mathrm{B}_1$/$(5a_1)^{-1}(2b_1)^{-1}(6a_1)^1$} \\
      Exp.        & \mc{4}{c}{19.63\cite{French_1988}} & \mc{4}{c}{} & \mc{4}{c}{} \\
      \cline{2-5} \cline{6-9} \cline{10-13}
      CC3         & 19.182 & 19.350 & 19.377 &     \x & 19.973 & 20.137 & 20.317 & 20.355 & 20.456 & 20.620 & 20.702 &    \x \\
      CCSDT       & 18.761 & 19.018 & 19.043 &     \x & 19.675 & 19.961 & 20.136 & 20.190 & 19.948 & 20.185 & 20.269 &    \x \\
      CC4         & 18.607 & 18.848 & 18.801 &     \x & 19.485 & 19.777 & 19.892 &     \x & 19.775 & 19.995 & 20.019 &    \x \\
      CCSDTQ      & 18.582 & 18.827 & 18.772 &     \x & 19.467 & 19.765 & 19.868 &     \x & 19.744 & 19.969 & 19.986 &    \x \\
      FCI         & 18.575 & 18.819 & 18.755 & 18.745 & 19.462 & 19.759 & 19.853 & 19.889 & 19.741 & 19.965 & 19.974 &    \x \\
      \hline
      Mol.        & \mc{4}{c}{\ce{PH3}} & \mc{8}{c}{\ce{HCl}} \\
      State/Conf. & \mc{4}{c}{$2~^{2}\mathrm{E}$/$(5a_1)^{-2}(4e_g)^{1}$} & \mc{4}{c}{$2~^{2}\Sigma^+$/$(2\pi)^{-2}(6\sigma)^{1}$} & \mc{4}{c}{$1~^{2}\Delta$/$(2\pi)^{-2}(6\sigma)^{1}$} \\
      Exp.        & \mc{4}{c}{} &  &  &  &  &  &  &  &  \\
      \cline{2-5} \cline{6-9} \cline{10-13}
      CC3         & 19.513 & 19.613 & 19.630 & 19.606 & 22.798 & 22.997 & 23.197 & 23.262 & 23.556 & 23.732 & 23.795 & 23.829 \\
      CCSDT       & 19.072 & 19.217 & 19.255 & 19.245 & 22.277 & 22.676 & 22.885 & 22.982 & 23.188 & 23.532 & 23.532 & 23.590 \\
      CC4         & 18.940 & 19.084 & 19.080 &     \x & 21.932 & 22.328 & 22.439 &     \x & 22.918 & 23.273 & 23.185 &     \x \\
      CCSDTQ      & 18.907 & 19.055 & 19.046 &     \x & 21.889 & 22.303 & 22.400 &     \x & 22.885 & 23.257 & 23.155 &     \x \\  
      FCI         & 18.897 & 19.047 & 19.025 & 18.997 & 21.878 & 22.293 & 22.377 & 22.463 & 22.881 & 23.254 & 23.142 & 23.185 \\
      \hline
      Mol.        & \mc{4}{c}{\ce{Ar}} & \mc{8}{c}{} \\
      State/Conf. & \mc{4}{c}{$2~^{2}\mathrm{P}$/$(3p)^{-2}(4s)^1$} & \mc{4}{c}{} & \mc{4}{c}{} \\
      Exp.        & \mc{4}{c}{34.21\cite{Kikas_1996}} &  &  &  &  &  &  &  &  \\
      \cline{2-5} \cline{6-9} \cline{10-13}
      CC3         & 31.956 & 32.384 & 32.638 & 32.719 &  &  &  &  &  &  &  &  \\
      CCSDT       & 32.381 & 32.811 & 33.220 & 33.423 &  &  &  &  &  &  &  &  \\
      CC4         & 31.913 & 32.397 & 32.650 & 32.833 &  &  &  &  &  &  &  &  \\
      CCSDTQ      & 31.933 & 32.420 & 32.678 & 32.869 &  &  &  &  &  &  &  &  \\  
      FCI         & 31.924 & 32.420 & 32.657 & 32.855 &  &  &  &  &  &  &  &  \\
    \end{tabular}
  \end{ruledtabular}
\end{table*}
\endgroup

\squeezetable
\begingroup
\begin{table*}
  \caption{Valence ionizations and satellite transition energies (in \si{\eV}) of the remaining molecules for various methods and basis sets. 
  AVXZ stands for aug-cc-pVXZ (where X = D, T, and Q). Selected experimental values are also reported.}
  \label{tab:tab9}
  \begin{ruledtabular}
    \begin{tabular}{rccccccccccccc}
      & \mc{4}{c}{Basis} & \mc{4}{c}{Basis} & \mc{4}{c}{Basis} \\
      \cline{2-5} \cline{6-9} \cline{10-13}
      Methods & \mcc{6-31$+$G$^{*}$} & \mcc{AVDZ} & \mcc{AVTZ} & \mcc{AVQZ} & \mcc{6-31$+$G$^{*}$} & \mcc{AVDZ} & \mcc{AVTZ} & \mcc{AVQZ} & \mcc{6-31$+$G$^{*}$} & \mcc{AVDZ} & \mcc{AVTZ} & \mcc{AVQZ}  \\
      \hline
      Mol.        & \mc{12}{c}{Carbon dioxide (\ce{CO2})} \\
      State/Conf. & \mc{4}{c}{$1~^{2}\Pi_g$/$(1\pi_g)^{-1}$} & \mc{4}{c}{$1~^{2}\Pi_u$/$(1\pi_u)^{-1}$} & \mc{4}{c}{$1~^{2}\Sigma_u^+$/$(3\sigma_u)^{-1}$} \\
      Exp.        & \mc{4}{c}{13.8 \cite{Tian_2012}} & \mc{4}{c}{17.6 \cite{Tian_2012}} & \mc{4}{c}{18.1 \cite{Tian_2012}} \\
      \cline{2-5} \cline{6-9} \cline{10-13}
      CC2         & 12.747 & 12.812 & 13.015 & 13.103 & 16.508 & 16.630 & 16.751\ph & 16.829\ph & 16.866\ph & 16.889\ph & 17.034\ph & 17.107 \\
      CCSD        & 13.490 & 13.561 & 13.787 & 13.879 & 17.767 & 17.809 & 17.996\ph & 18.080\ph & 17.885\ph & 18.006\ph & 18.171\ph & 18.259 \\
      CC3         & 13.456 & 13.528 & 13.696 & 13.773 & 17.275 & 17.335 & 17.463\ph & 17.531\ph & 17.708\ph & 17.829\ph & 17.928\ph & 17.999 \\
      CCSDT       & 13.474 & 13.539 & 13.716 & 13.794 & 17.351 & 17.406 & 17.553\ph & 17.628\ph & 17.708\ph & 17.821\ph & 17.930\ph & 18.005 \\
      CC4         & 13.488 & 13.562 & 13.733 &     \x & 17.339 & 17.396 & 17.521\ph &     \x\ph & 17.783\ph & 17.904\ph & 18.009\ph &     \x \\
      CCSDTQ      & 13.491 & 13.563 & 13.734 &     \x & 17.341 & 17.396 & 17.522\ph &     \x\ph & 17.750\ph & 17.869\ph & 17.972\ph &     \x \\
      FCI         & 13.496 & 13.567 & 13.733 & 13.823 & 17.337 & 17.391 & 17.513(7) & 17.618(6) & 17.766(1) & 17.889(1) & 17.996(2) &     \x \\
      $G_0W_0$    & 13.743 & 13.837 & 14.221 & 14.425 & 18.184 & 18.209 & 18.519\ph & 18.697\ph & 18.335\ph & 18.474\ph & 18.791\ph & 18.994 \\
      qs$GW$      & 13.712 & 13.806 & 14.054 & 14.182 & 17.950 & 18.008 & 18.200\ph & 18.315\ph & 18.120\ph & 18.264\ph & 18.442\ph & 18.566 \\
      G$_0$F(2)   & 12.783 & 12.854 & 13.129 & 13.254 & 16.993 & 17.012 & 17.224\ph & 17.332\ph & 16.714\ph & 16.840\ph & 17.035\ph & 17.153 \\
      $G_0T_0$    & 13.284 & 13.334 & 13.586 &     \x & 17.700 & 17.696 & 17.890\ph &     \x\ph & 17.693\ph & 17.816\ph & 17.995\ph &     \x \\
      \hline
      Mol.        & \mc{4}{c}{Carbon dioxide (\ce{CO2})} & \mc{8}{c}{} \\
      State/Conf. & \mc{4}{c}{$1~^{2}\Sigma_g^+$/$(4\sigma_g)^{-1}$} & \mc{4}{c}{} & \mc{4}{c}{} \\
      Exp.        & \mc{4}{c}{19.4 \cite{Tian_2012}} & \mc{4}{c}{} & \mc{4}{c}{} \\
      \cline{2-5} \cline{6-9} \cline{10-13}
      CC2         & 17.746\ph & 17.864\ph & 17.980\ph & 18.048 &  &  &  &  &  &  &  &  \\
      CCSD        & 19.237\ph & 19.351\ph & 19.523\ph & 19.605 &  &  &  &  &  &  &  &  \\
      CC3         & 18.959\ph & 19.073\ph & 19.184\ph & 19.250 &  &  &  &  &  &  &  &  \\
      CCSDT       & 18.968\ph & 19.075\ph & 19.198\ph & 19.268 &  &  &  &  &  &  &  &  \\
      CC4         & 19.047\ph & 19.158\ph & 19.274\ph &     \x &  &  &  &  &  &  &  &  \\
      CCSDTQ      & 19.008\ph & 19.119\ph & 19.232\ph &     \x &  &  &  &  &  &  &  &  \\
      FCI         & 19.024(1) & 19.137(2) & 19.254(3) &     \x &  &  &  &  &  &  &  &  \\
      $G_0W_0$    & 19.730\ph & 19.856\ph & 20.150\ph & 20.334 &  &  &  &  &  &  &  &  \\
      qs$GW$      & 19.474\ph & 19.604\ph & 19.776\ph & 19.888 &  &  &  &  &  &  &  &  \\
      G$_0$F(2)   & 17.949\ph & 18.069\ph & 18.245\ph & 18.348 &  &  &  &  &  &  &  &  \\
      $G_0T_0$    & 19.083\ph & 19.193\ph & 19.353\ph &     \x &  &  &  &  &  &  &  &  \\
      \hline
      Mol.        & \mc{12}{c}{Formaldehyde (\ce{CH2O})} \\
      State/Conf. & \mc{4}{c}{$1~^{2}\mathrm{B}_2 $/$(2b_2)^{-1}$} & \mc{4}{c}{$1~^{2}\mathrm{B}_1 $/$(1b_1)^{-1}$} & \mc{4}{c}{$1~^{2}\mathrm{A}_1 $/$(5a_1)^{-1}$} \\
      Exp.        & \mc{4}{c}{10.9 \cite{vonNiessen_1980}} & \mc{4}{c}{14.5 \cite{vonNiessen_1980}} & \mc{4}{c}{16.1 \cite{vonNiessen_1980}} \\
      \cline{2-5} \cline{6-9} \cline{10-13}
      CC2         & \z9.444 & \z9.553 & \z9.753 & \z9.828 & 13.854 & 13.903 & 14.053\ph & 14.127\ph & 14.596 & 14.705 & 14.820\ph & 14.893\ph \\
      CCSD        &  10.486 &  10.625 &  10.848 &  10.924 & 14.410 & 14.480 & 14.606\ph & 14.667\ph & 15.833 & 15.964 & 16.105\ph & 16.181\ph \\
      CC3         &  10.555 &  10.710 &  10.873 &  10.932 & 14.377 & 14.482 & 14.562\ph & 14.609\ph & 15.832 & 15.985 & 16.063\ph & 16.119\ph \\
      CCSDT       &  10.533 &  10.680 &  10.840 &  10.898 & 14.381 & 14.478 & 14.562\ph & 14.609\ph & 15.811 & 15.953 & 16.030\ph & 16.086\ph \\
      CC4         &  10.578 &  10.736 &  10.897 &      \x & 14.395 & 14.504 & 14.578\ph &     \x\ph & 15.879 & 16.039 & 16.114\ph &     \x\ph \\
      CCSDTQ      &  10.574 &  10.729 &  10.887 &      \x & 14.391 & 14.498 & 14.572\ph &     \x\ph & 15.864 & 16.018 & 16.090\ph &     \x\ph \\
      FCI         &  10.582 &  10.739 &  10.899 &  10.954 & 14.395 & 14.502 & 14.578(1) & 14.618(9) & 15.876 & 16.032 & 16.106(1) & 16.161(4) \\
      $G_0W_0$    &  10.835 &  10.996 &  11.380 &  11.556 & 14.232 & 14.283 & 14.587\ph & 14.759\ph & 16.170 & 16.292 & 16.602\ph & 16.794\ph \\
      qs$GW$      &  10.780 &  10.990 &  11.241 &  11.342 & 14.395 & 14.543 & 14.693\ph & 14.780\ph & 16.118 & 16.313 & 16.459\ph & 16.557\ph \\
      G$_0$F(2)   & \z9.605 & \z9.712 & \z9.981 &  10.088 & 13.740 & 13.772 & 13.985\ph & 14.093\ph & 14.747 & 14.856 & 15.040\ph & 15.153\ph \\
      $G_0T_0$    &  10.345 &  10.407 &  10.655 &      \x & 13.879 & 13.896 & 14.082\ph &     \x\ph & 15.625 & 15.716 & 15.888\ph &     \x\ph \\
      \hline
      Mol.        & \mc{4}{c}{Formaldehyde (\ce{CH2O})} & \mc{8}{c}{Borane (\ce{BH3})} \\
      State/Conf. & \mc{4}{c}{$2~^{2}\mathrm{B}_2 $/$(1b_2)^{-1}$} & \mc{4}{c}{$1~^{2}\mathrm{E}' $/$(1e')^{-1}$} & \mc{4}{c}{$1~^{2}\mathrm{A}_1' $/$(2a_1')^{-1}$} \\
      Exp.        & \mc{4}{c}{17.0 \cite{vonNiessen_1980}} & \mc{4}{c}{} & \mc{4}{c}{} \\
      \cline{2-5} \cline{6-9} \cline{10-13}
      CC2         & 16.704 & 16.738 & 16.866 & 16.921\ph & 12.942 & 13.081 & 13.231 & 13.276 & 18.210 & 18.337 & 18.427 & 18.465 \\
      CCSD        & 17.256 & 17.338 & 17.494 & 17.553\ph & 12.995 & 13.204 & 13.342 & 13.375 & 18.100 & 18.300 & 18.398 & 18.428 \\
      CC3         & 16.898 & 16.992 & 17.102 & 17.148\ph & 12.941 & 13.191 & 13.309 & 13.334 & 18.018 & 18.249 & 18.326 & 18.349 \\
      CCSDT       & 16.938 & 17.036 & 17.156 & 17.207\ph & 12.938 & 13.189 & 13.304 & 13.330 & 18.002 & 18.232 & 18.307 & 18.331 \\
      CC4         & 16.905 & 17.005 & 17.107 &     \x\ph & 12.938 & 13.193 & 13.306 & 13.331 & 18.001 & 18.236 & 18.308 & 18.331 \\
      CCSDTQ      & 16.906 & 17.005 & 17.108 &     \x\ph & 12.938 & 13.193 & 13.307 &     \x & 18.001 & 18.237 & 18.309 &     \x \\
      FCI         & 16.901 & 17.002 & 17.107 & 17.156(5) & 12.938 & 13.194 & 13.307 & 13.332 & 18.001 & 18.238 & 18.310 & 18.332 \\
      $G_0W_0$    & 17.651 & 17.714 & 18.015 & 18.162\ph & 13.202 & 13.395 & 13.678 & 13.780 & 18.320 & 18.470 & 18.685 & 18.780 \\
      qs$GW$      & 17.463 & 17.635 & 17.821 & 17.909\ph & 13.047 & 13.371 & 13.584 & 13.651 & 18.111 & 18.410 & 18.544 & 18.601 \\
      G$_0$F(2)   & 16.460 & 16.463 & 16.656 & 16.742\ph & 12.989 & 13.093 & 13.278 & 13.341 & 18.255 & 18.342 & 18.459 & 18.515 \\
      $G_0T_0$    & 17.317 & 17.298 & 17.473 &     \x\ph & 13.130 & 13.179 & 13.327 &     \x & 18.536 & 18.598 & 18.683 &     \x \\
      \hline
      Mol.        & \mc{8}{c}{Formaldehyde (\ce{CH2O})} & \mc{4}{c}{} \\
      State/Conf. & \mc{4}{c}{$2~^{2}\mathrm{B}_1 $/$(2b_2)^{-1}(2b_1)^{1}$} & \mc{4}{c}{$3~^{2}\mathrm{B}_2 $/$(1b_1)^{-1}(2b_2)^{-1}(2b_1)^{1}$} & \mc{4}{c}{} \\
      Exp.        & \mc{4}{c}{} & \mc{4}{c}{} & \mc{4}{c}{} \\
      \cline{2-5} \cline{6-9} 
      CC3         & 16.564 & 16.632 & 16.800\ph & 16.847 & 18.826 & 18.991 & 19.154\ph & 19.208 &  &  &  &  \\
      CCSDT       & 16.684 & 16.786 & 17.004\ph & 17.083 & 18.761 & 18.956 & 19.153\ph & 19.233 &  &  &  &  \\
      CC4         & 16.348 & 16.407 & 16.495\ph &     \x & 18.503 & 18.663 & 18.751\ph &     \x &  &  &  &  \\
      CCSDTQ      & 16.363 & 16.425 & 16.522\ph &     \x & 18.515 & 18.679 & 18.772\ph &     \x &  &  &  &  \\
      FCI         & 16.345 & 16.409 & 16.498(1) &     \x & 18.512 & 18.678 & 18.763(1) &     \x &  &  &  &  \\
      \hline
      Mol.        & \mc{8}{c}{Borane (\ce{BH3})} & \mc{4}{c}{ } \\
      State/Conf. & \mc{4}{c}{$1~^{2}\mathrm{B}_1' $/$(1e')^{-2}(1b_1)^1$} & \mc{4}{c}{$2~^{2}\mathrm{E}' $/$(1e')^{-2}(1b_1)^1$} & \mc{4}{c}{} \\
      Exp.        & \mc{4}{c}{} & \mc{4}{c}{} & \mc{4}{c}{} \\
      \cline{2-5} \cline{6-9} 
      CC3         & 19.485 & 19.803 & 19.889 & 19.894 & 20.026 & 20.212 & 20.324 & 20.346 &  &  &  &  \\
      CCSDT       & 18.896 & 19.228 & 19.331 & 19.355 & 19.482 & 19.669 & 19.789 & 19.824 &  &  &  &  \\
      CC4         & 18.790 & 19.130 & 19.209 & 19.226 & 19.403 & 19.602 & 19.704 & 19.732 &  &  &  &  \\
      CCSDTQ      & 18.756 & 19.100 & 19.178 &     \x & 19.379 & 19.584 & 19.684 &     \x &  &  &  &  \\
      FCI         & 18.754 & 19.099 & 19.176 & 19.194 & 19.377 & 19.583 & 19.685 & 19.714 &  &  &  &  \\
      \end{tabular}
  \end{ruledtabular}
\end{table*}
\endgroup

\section{Conclusion}
\label{sec:conclusion}

We have reported 42 FCI satellite transition energies computed in 23 small molecules.
These energies have been calculated with increasingly large basis sets ranging from Pople's 6-31+G* to Dunning's aug-cc-pVXZ (where X = D, T, and Q).
In addition, 58 FCI reference values for outer- and inner-valence IPs of the same molecular set have been presented.
This work is the tenth layer of reference values of the \textsc{quest} database \cite{Loos_2020d,Veril_2021} and the first one to include charged excitations.

Various CC methods have been employed to compute IPs (CC2, CCSD, CC3, CCSDT, CC4, and CCSDTQ) and satellite transition energies (CC3, CCSDT, CC4, and CCSDTQ), and their performances have been assessed using the FCI reference values.
It has been shown that CC3 and CC4 are faithful approximations of CCSDT and CCSDTQ for IPs, respectively, while the CC2 approximate treatment of double excitations induces large errors with respect to CCSD.
For the satellites, our study reveals that chemical accuracy is reached only at the CCSDTQ level, highlighting the intricate and complex correlation effects involved in such states and their overall challenging nature for computational methods.

The performance of various propagator methods ($G_0W_0$, G$_0$F(2), $G_0T_0$, and qs$GW$) have also been gauged.
The poor performance of these methods for satellite transition energies has been discussed in detail.
These results call for the development of new methods capable of describing such states.
For example, considering explicitly the three-body Green's function in order to describe IPs and satellites on an equal footing could offer significant advantages. \cite{Riva_2022,Riva_2023}
Studying the convergence of the ADC hierarchy using these new benchmark values is another possible outlook.
Finally, assessing methods designed in the condensed matter community, such as the cumulant Green's function, on these small molecular systems would certainly be interesting. Work along this line is presently underway. \cite{Loos_2024a}

One obvious perspective that needs to be addressed is the extension to transition intensities, which are of crucial importance for direct comparisons with experimental spectra. An approximate electronic structure method should not only aim to accurately describe the excited-state energy but also the transition intensities associated with it. 
Within the present SCI formalism, computing intensities is not straightforward but this is feasible, as demonstrated in Refs.~\onlinecite{Mejuto-Zaera_2021,Ferte_2020, Ferte_2022}, and is planned for future investigation.

\acknowledgements{
This project has received funding from the European Research Council (ERC) under the European Union's Horizon 2020 research and innovation programme (Grant agreement No.~863481).
This work used the HPC resources from CALMIP (Toulouse) under allocations 2023-18005 and 2024-18005.
The authors thank Abdallah Ammar, F\'abris Kossoski, Yann Damour, Alexander Sokolov, Devin Matthews, Anthony Scemama, and Denis Jacquemin for helpful comments and/or insightful discussions.}

\section*{Associated Content}

The \SupInf includes the geometry of the 23 molecules considered in this study as well as a json file for each molecule.
This json file contains all the IPs and satellite transition energies of a given molecule as well as the FCI incertitudes and the spectral weight associated with the $G_0W_0$, G$_0$F(2), and $G_0T_0$ quasiparticle energies.


\section*{References}

\bibliography{shakeup}

\end{document}